\documentclass{article}

\usepackage[utf8]{inputenc}
\usepackage[english]{babel}
\usepackage{newunicodechar}
\usepackage{float} 
\usepackage{caption}  

\newunicodechar{μ}{\ensuremath{\mu}}      
\newunicodechar{₂}{\textsubscript{2}}     
\newunicodechar{⁻}{\textsuperscript{-}}   
\newunicodechar{，}{,}                    
\newunicodechar{∼}{\ensuremath{\sim}}     
\newunicodechar{ω}{\ensuremath{\omega}}
\newunicodechar{ν}{\ensuremath{\nu}}
\newunicodechar{γ}{\ensuremath{\gamma}}
\newunicodechar{α}{\ensuremath{\alpha}}
\newunicodechar{δ}{\ensuremath{\delta}}
\newunicodechar{Φ}{\ensuremath{\Phi}}
\newunicodechar{λ}{\ensuremath{\lambda}}
\newunicodechar{ω}{\ensuremath{\omega}}
\newunicodechar{ν}{\ensuremath{\nu}}
\newunicodechar{γ}{\ensuremath{\gamma}}
\newunicodechar{≈}{\ensuremath{\approx}}
\newunicodechar{−}{-} 

\usepackage[letterpaper,top=2cm,bottom=2cm,left=3cm,right=3cm,marginparwidth=1.75cm]{geometry}

\usepackage{amsmath}
\usepackage{graphicx}
\usepackage[colorlinks=true, allcolors=blue]{hyperref}

\title{Photocatalytic CO\textsubscript{2} Reduction Enhanced by Synergetic Interactions among Photon，Phonon, and Molecule}
\author{Chen Sun \textsuperscript{a }, Yimin Xuan\textsuperscript{ a} *}

\begin{document}
\maketitle

\begin{abstract}
Photocatalytic CO$_2$ reduction is limited by inefficient CO$_2$ activation and poor solar spectrum utilization. Here, we discovered and revealed the vibration coupling mechanism among photons, phonons, and molecules, which remarkably enhances the photocatalytic catalysis of CO$_2$ into fuels. We designed the nitrogen-doped Cu$_2$O-based catalyst loaded onto the quartz optical substrate. N-doping Cu$_2$O converts linear geometry of adsorbed CO₂ molecules, which efficiently lowers the activation barrier and facilitates the dissociation of CO$_2$. Once the Cu-based catalyst is combined with a micro-pillar quartz film, the system induces vibrational strong coupling (VSC) between the asymmetric CO$_2$ stretching mode and surface phonon polariton resonances. These resonances arise from the photothermal conversion of incident solar photons on the micro-pillars. The resonant coupling phenomena were further verified by Fourier-transform infrared spectroscopy using Synchrotron Radiation Source (SRS), which directly confirmed the interactions between molecular vibrations and photonic–phononic modes. The synergetic functions originated from this hybrid architecture achieves a CO yield of 167.7 μmol·h\textsuperscript{-1}·g\textsuperscript{-1} under pure water conditions, which is the highest reported yield for Cu$_2$O-based photocatalysts with 46\% enhancement over non-VSC systems. This work uncovers a novel photo-thermal mechanism. It further provides a new strategy to control bond activation in photocatalytic CO$_2$ conversion through light–vibration–matter coupling.
\end{abstract}

\section{Introduction}

The increasing concentration of atmospheric CO₂ driven by anthropogenic activities has been unequivocally linked to global climate destabilization through the greenhouse effect\textsuperscript{1-3}. Photocatalytic CO₂ reduction (CO₂RR), first proposed in 1972\textsuperscript{4}, offers a dual promise of mitigating carbon emissions and synthesizing renewable fuels\textsuperscript{5-7}. However, practical implementation remains constrained by fundamental kinetic limitations: (1) The exceptionally stable C=O bond (∼750 kJ mol⁻¹ dissociation energy)\textsuperscript{8}, (2) the instability of transient metal-C/O intermediates relative to adsorbed CO₂\textsuperscript{9}, and (3) inefficient utilization of solar spectra by conventional photocatalysts\textsuperscript{10}. Traditional strategies including heteroatom doping, vacancy engineering, and facet modulation have been explored to asymmetrically polarize CO₂ molecules and strengthen adsorption\textsuperscript{7,11,12}. For example, N-doped Cu₂O can reach adsorption energies as high as –1.6 eV. While such approaches partially improve CO₂ activation, they are fundamentally restricted by competitive hydrogen evolution reactions (HER), inefficient charge-carrier separation, and narrow light-harvesting ranges. Thus, a more precise and energy-selective pathway is required to regulate the activation and dissociation of CO₂ molecules during photocatalysis.

An efficient solution to above problems is to break the symmetrical line structure of CO\textsubscript{2} molecule by reconstructing the active center on catalysts, such as doping other elements and making vacancies. In this way, it can perturb the electronic structures of nonpolar CO\textsubscript{2} and polarize the molecular structure, thereby increasing the adsorption energy of CO\textsubscript{2} molecules and accelerating the kinetic process of CO\textsubscript{2} activation. However, the efficiency of CO\textsubscript{2}RR is still limited due to factors such as the competitive hydrogen evolution reaction (HER), charge carrier separation, and spectral utilization. Therefore, further steps are needed for the precise regulation of the adsorbed CO\textsubscript{2} molecules in the dissociation kinetic processes of photocatalytic CO\textsubscript{2}RR. In the photocatalytic process of multi-sites adsorbed CO\textsubscript{2}, it can be observed that the C=O bond undergoes stretching, making it more susceptible to breaking, which is directly related to the stretching vibrational mode of the C=O bond. Indeed, many efforts have been many attempts to achieve precise regulation of chemical reactions through vibrational modes\textsuperscript{13,14}. The vibrational strong coupling (VSC) has been shown to accelerate or hinder molecular chemical reactivity, and successfully realized with photonic (e.g., Fabry–Perot) micro-cavity modes\textsuperscript{15}, surface plasmon polaritons (SPPs) of thin metallic films or gratings, and surface phonon polaritons (SPhPs) of dielectric polar crystals\textsuperscript{16}. However, there are currently few applications that directly regulate catalytic reactions through VSC, due to factors such as low quality factors Q and dark states. Enlightened by the above premises，we envision that promoting photocatalytic reactions through the construction of vibrational strong coupling (VSC) among phonon, molecule, and photon. 

Herein, we designed and fabricated quartz plates etched with a micro-pillar to induce the mid-infrared localized plasmons, and conducted experimental studies for VSC between phonon and CO\textsubscript{2} molecules by loading the photocatalyst. Firstly, we analyzed the adsorption properties and kinetic processes of CO\textsubscript{2} molecules on the Cu\textsubscript{2}O(100) surface and proposed an improved catalyst design scheme. The Cu\textsubscript{2}O is a good candidate for the reduction of CO\textsubscript{2}\textsuperscript{17,18}. However, the stable crystal plane (100) of Cu\textsubscript{2}O, which is its primary exposed facet, has a relatively poor adsorption capacity for carbon dioxide, thereby limiting its photocatalytic activity\textsuperscript{19}. In our calculations, it shows that N doping enhances the adsorption capability, with a surface adsorption energy of -1.6eV. By calculating the adsorption states and dissociation processes of carbon dioxide on the N-Cu\textsubscript{2}O surface, we identified the asymmetric stretching vibration mode at 1136.8 cm\textsuperscript{-1}(1136.2cm\textsuperscript{-1} in experimental measurement) as an important vibrational mode in the activation and dissociation of carbon dioxide. And the frequency of C=O vibration is in the Reststrahlen band of quartz:1072cm\textsuperscript{-1}\~1215cm\textsuperscript{-1} \textsuperscript{20,21}. Therefore, we designed the quartz films as the SPhP resonators by etching micro-pillararrary in its surface. By controlling the dimensions of the micro-pillars, we fabricated different quartz sheets to tune the SPhP modes. Additionally, considering carrier separation and light absorption effects, iron (Fe) and silver (Ag) were incorporated. Furthermore, by loading N-Cu\textsubscript{2}O/Ag/FeO\textsubscript{x} on quartz films with micro-pillars (QF-MP), we achieved photocatalytic yields of 167.7 μmol/h/g for CO and 57.3 mol/h/g for CH\textsubscript{4}, and confirmed the occurrence of VSC effects by far-field observation. Combining theoretical calculations, we elucidated the mechanism for enhancing the reaction rate is shown in the Figure 1, when the specific C=O vibrational states of the adsorbed CO\textsubscript{2} molecules couple with the SPhP, the energy of that vibrational mode is enhanced. This increases the overall energy of the adsorbed CO\textsubscript{2} molecules, thereby lowering the energy barrier required for the catalytic reaction. This work establishes a new framework for leveraging phonon–photon–molecule coupling to modulate surface reaction pathways, providing a generalizable strategy for efficient photocatalytic CO₂ conversion.
\begin{figure}[H]
    \centering
    \includegraphics[width=1\linewidth]{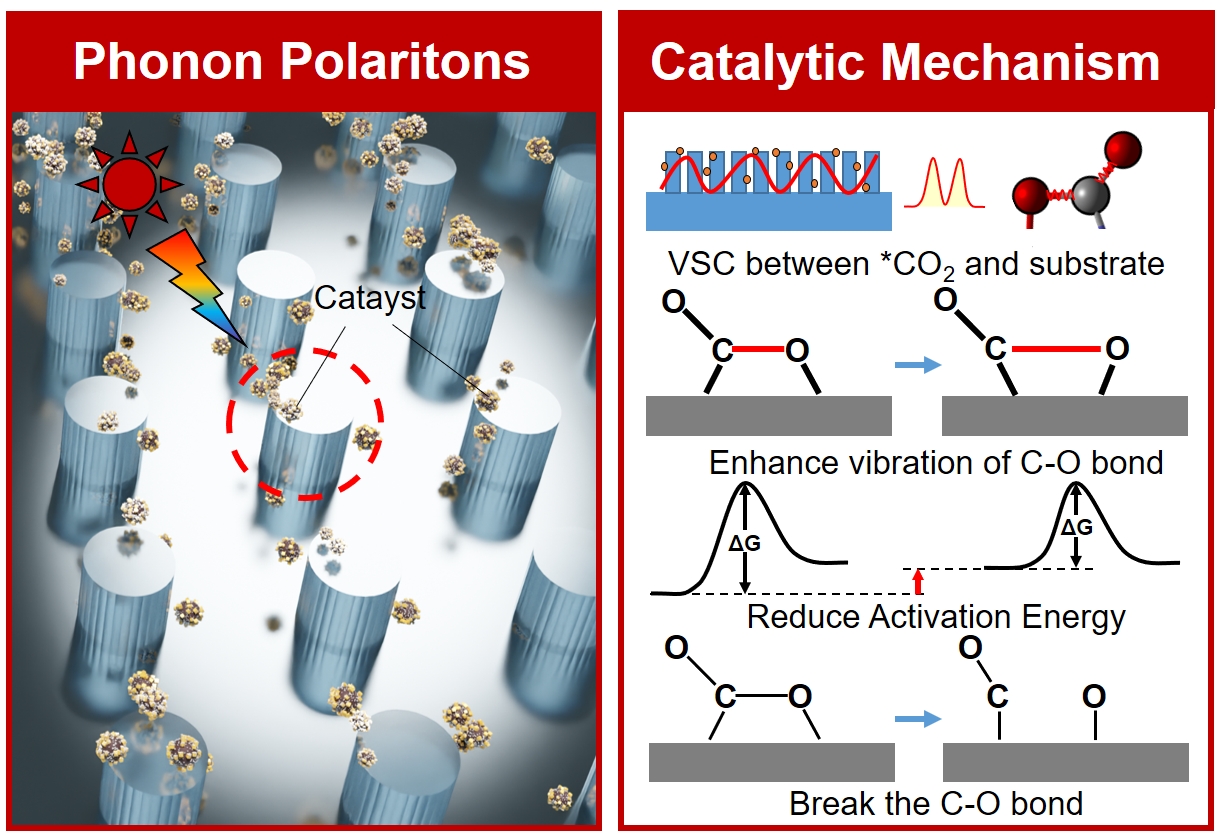}
    Figure 1. Schematic diagram of the mechanism of promoting photocatalytic CO\textsubscript{2} reduction reaction by loading N-CAF catalysts on the quartz film etched micro-pillar arrays to realize of VSC between CO\textsubscript{2} molecules and surface phonons.
\end{figure}

\section{Results and Discussion}

\subsection{Characterization and Tuning of CO$_2$ in Adsorbed States}

The gaseous CO\textsubscript{2} molecules are symmetrical linear molecules, making them difficult to decompose (vibrational modes shown in the Figure S1). In contrast, CO\textsubscript{2} molecules are more easily dissociated when adsorbed on the catalyst surface in a bent configuration. To precisely regulate the reaction behavior of CO₂ molecules adsorbed on the catalyst surface, constructing a well-defined adsorption configuration is essential. Considering that the adsorption state of CO₂ strongly depends on the exposed crystal facets of the catalyst, we selected a material with a single exposed facet. Furthermore, the chosen facet itself is required to be chemically inert, so that a uniform CO₂ adsorption geometry can be achieved through the deliberate introduction of a single active site. The cubic-shaped cuprous oxide (Cu₂O) nanoparticles not only fulfill the above requirements but also have a suitable bandgap (\~1.9eV) for photocatalytic applications. As shown in Fig 2(a) ,the exposed surface is the (100) facet, while the (110) plane exists only along the edges of the cube\textsuperscript{19}. The (100) facet of Cu\textsubscript{2}O serves as a stable crystal face, where only exhibits stable physical adsorption states. As shown in Fig. 2(b–c), nitrogen (N) doping did not alter the morphology of the catalyst. And the results of X-ray diffraction (XRD) characterization indicate no change in the crystal phase(Fig. 2d). X-ray photoelectron spectroscopy (XPS) analysis further demonstrated that nitrogen atoms were successfully incorporated into the surface of Cu₂O and formed chemical bonds with both Cu and O elements(Fig 2e). As shown in Figure 2f, nitrogen atoms bond with carbon atoms of CO\textsubscript{2} on the N-Cu\textsubscript{2}O (100) surface resulting in a bent adsorption structure, which facilitates the activation and dissociation of CO\textsubscript{2} molecules (as illustrated in Figure 1). Furthermore, we calculated the vibration modes of CO\textsubscript{2} molecules adsorbed on the N-Cu\textsubscript{2}O (100) surface, as shown in Figure 2g. The main vibrational modes include the stretching vibration ν\textsubscript{1} of C-O1 in the molecular plane, the asymmetric stretching vibration ν\textsubscript{2} of CO\textsubscript{2}. Notably, the calculated vibrational frequency of the ν\textsubscript{2} mode is 1136.8 cm⁻¹, which is directly related to the CO\textsubscript{2} dissociation process. Furthermore, we characterized the adsorbed CO₂ on the surfaces of Cu₂O and N-doped Cu₂O nanoparticles under an Ar atmosphere using in-situ DRIFTS. As shown in Fig 2h, while N-doped Cu₂O showed a strong C–O vibration at 1036 cm⁻¹, the unmodified sample exhibited only a negligible peak near 1064 cm⁻¹, possibly from edge-adsorbed CO₂ on the edge (111) facets."
\begin{figure}[H]
    \centering
    \includegraphics[width=1\linewidth]{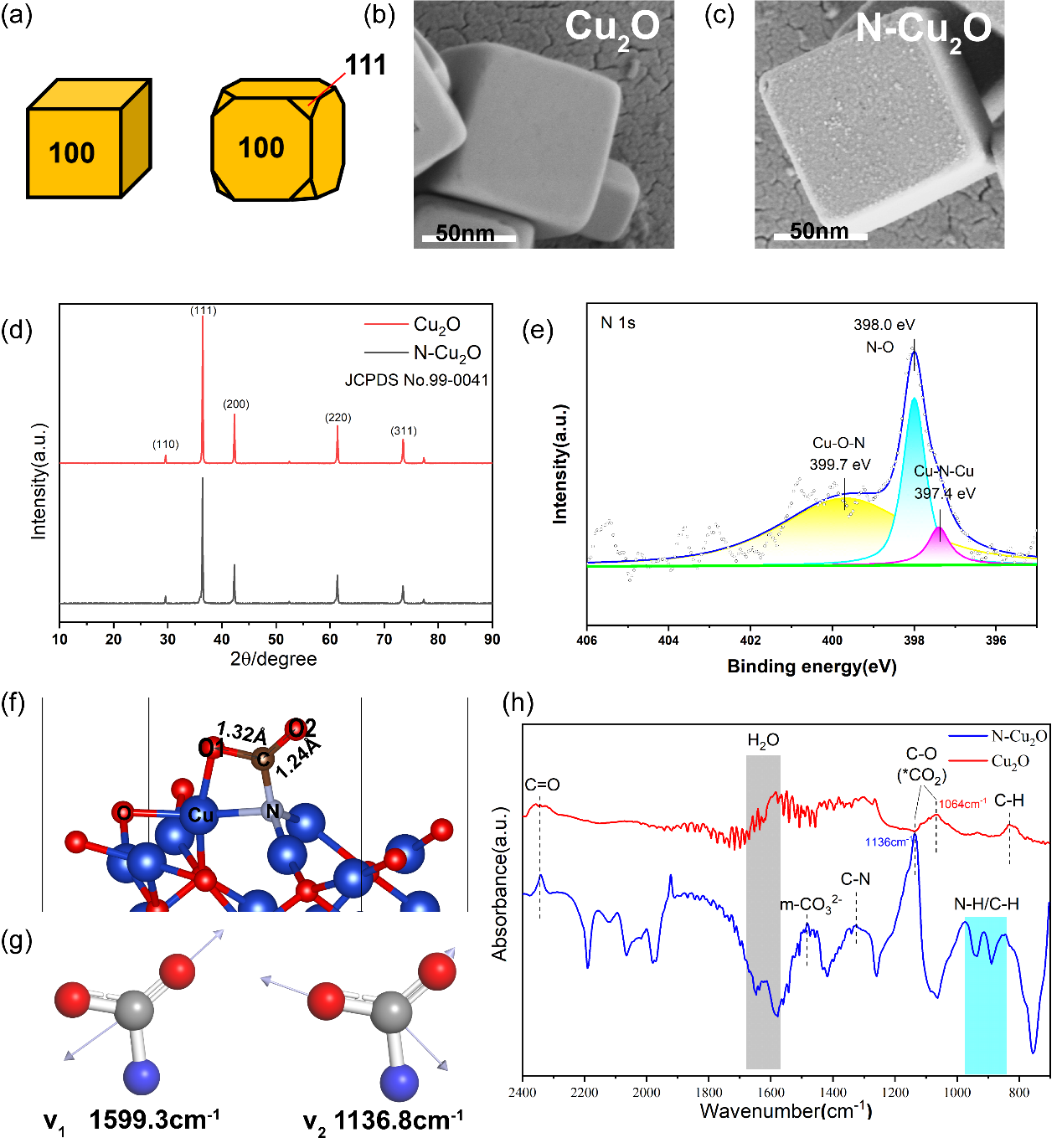}
Figure 2. (a) The morphology of cubic cuprous oxide (Cu\textsubscript{2}O) nanoparticles and the corresponding crystal facets ;(b),(c) Scanning electron microscopy images of Cu\textsubscript{2}O nanoparticle and N doped Cu\textsubscript{2}O nanoparticle;(d) XRD patterns of Cu\textsubscript{2}O and N-Cu\textsubscript{2}O (e) XPS spectra of N-CAF Cu 2p (f) the configuration of CO\textsubscript{2} molecules adsorbed on the N-Cu\textsubscript{2}O surface (g) the main vibrational modes of CO\textsubscript{2} in (f); (h) In-situ DRIFTS measurements of Cu\textsubscript{2}O and Cu\textsubscript{2}O nanoparticles with CO\textsubscript{2} adsorbed.
\end{figure}

\subsection{The Design and Characterization of the VSC Substrate}

To enhance the v\textsubscript{2} vibrational mode and facilitate the dissociation of CO\textsubscript{2} molecules on the catalyst surface, we designed the substrates made of etched micro-column arrays of α-quartz, as shown in Figure 3a. The optical properties of plasmonic materials were simulated by COMSOL\textsuperscript{®} software package (details in S1.3).Since the position of the SPhP resonance peak is related to the size and sharp of the pillars (Figure 3b), we simulated the electric field distribution and optical field around quartz micro-columns with different radii r\textsubscript{q} ranging from 0.8 μm to 1.5 μm. As shown in Figure 3c,d. The results showed that the SPhP modes are primarily located at both sides of the tops part of the micro-pillars, with a field enhancement ≈10². As shown in Figure 2f, the resonance peaks of the reflection spectrum redshift from 1133 cm⁻¹ to 1143 cm⁻¹ as the r\textsubscript{q} value decreases, which is consistent with the experimentally measured spectrum (as shown in Figure 4h). Notably, both experimental and numerical simulations exhibit a significant "pinning" effect due to the near-zero refractive index of the quartz substrate at the resonant frequencies (as shown in Figure 5a), indicating a prominent sign of plasma modes on the ENZ substrate. The Q factors of these plasma modes range from 20 (rq = 0.8 μm) to 70 (rq = 1.5 μm), which is very high for mid-infrared plasmonic structures and approaches the Q () factors of molecular vibrations. The sharp plasmonic resonances on the quartz substrate are used to interact with CO\textsubscript{2} molecules, which have a molecular vibration mode of C-O asymmetric stretching (approximately 1136 cm⁻¹), close to the resonance peaks of these plasmonic modes (as shown in Table S1).
\begin{figure}[H]
    \centering
    \includegraphics[width=1\linewidth]{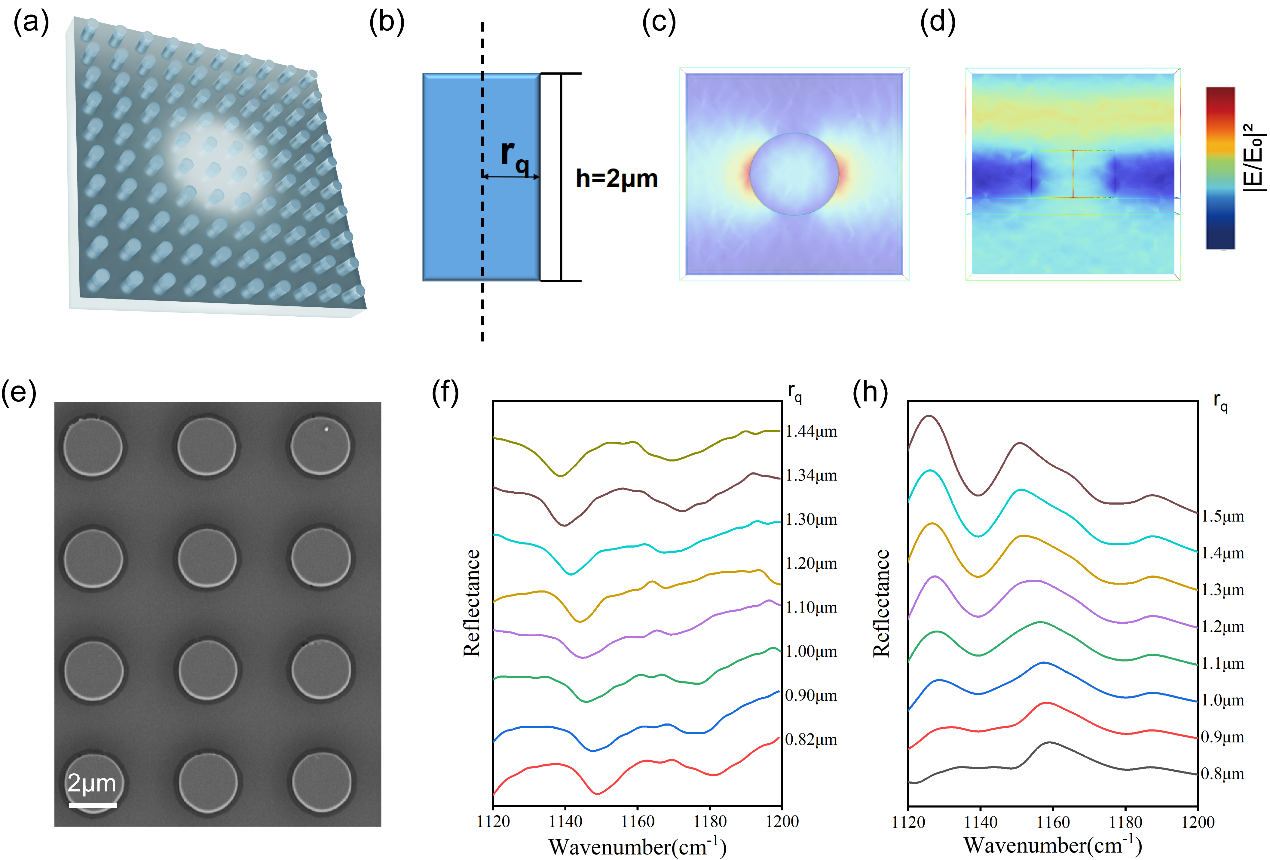}
Figure 3. (a) Schematics of the micro quartz-pillars array etched on quartz sheet. The quartz-pillars form an array with an inter-gap of 2.6 μm along both axes. (b) Schematic of micro quartz-pillar with a 2.0μm height and their radius r ranges from 0.8 to 1.44μm.; (e) Optical micrograph of QF-MP (r\textsubscript{q} = 1.1 μm). The scale bar indicates 2 μm; (f) Experimentally measured reflectance of QF-MP of different r\textsubscript{q}.;(h) Simulated length-dependent reflectance of quartz-pillars arrays. The total radius r ranges from 0.8 to 1.44μm.
\end{figure}
By loading N-Cu\textsubscript{2}O onto these QF-MP substrates and introducing a CO\textsubscript{2} atmosphere until the CO\textsubscript{2} molecules were fully adsorbed. The reflected spectra showed two resonance peaks on each side of ω\textsubscript{m}, indicating mode splitting for each diameter of the quartz micropillars (as shown in Figure 5b). Moreover, as the diameter varied, the resonance peaks shifted accordingly, exhibiting an anti-crossing behavior relative to ω\textsubscript{m} (as shown in Figure 5c). In order to study the coupling strength of phonon-molecule resonance, all vibration modes were modeled as classical harmonic oscillators, as shown in S1.5. The resonant frequencies (resonant damping) of phonon polaritons and molecule vibration ν\textsubscript{2} of CO\textsubscript{2} molecules are defined by ω\textsubscript{p} and ω\textsubscript{m} (γ\textsubscript{p} and γ\textsubscript{m}), respectively. The interaction between them is described by the coupling strength g. Based on the modelled fittings (Fig. 5b), the values of  ω\textsubscript{m}, ω\textsubscript{m}, γ\textsubscript{p}, γ\textsubscript{m} and g for each r\textsubscript{q} were obtained (detailed in Table S3). Using the above values, calculate the resonant frequency ω\textsubscript{±} \textsuperscript{22}:
\begin{equation}
    \omega_{\pm} = \frac{1}{2} (\omega_p + \omega_m) \pm \frac{1}{2} Re \left[ \sqrt{4|g|^2 + \left[\delta + i \left(\frac{\gamma_m}{2} - \frac{\gamma_p}{2}\right)\right]^2} \right]
    \label{eq:omega_pm}
    \tag{1}
\end{equation}
where δ( =ω\textsubscript{p} - ω\textsubscript{m}) is the detuning between the SPhP resonance and the CO\textsubscript{2} molecular vibration. The hybrid-mode frequencies ω\textsubscript{+} and ω\textsubscript{− }presents an anti-crossing feature in Fig 5c, that satisfies the necessary conditions for strong coupling resonance. Furthermore, we plots g as a function of r\textsubscript{q} in Fig 5d, showing an average g of 7.3 cm\textsuperscript{−1} which is smaller than γ\textsubscript{p }or γ\textsubscript{m}. It is the signal that the couplings of the most samples are not vary strong. To quantify the strong coupling phenomenon, the criterion C, defined by Equation (2)\textsuperscript{22}, which average is about 0.2.
\begin{equation}
   C = \frac{|g|}{|\gamma_m + \gamma_p|}
   \label{eq:normalized_coefficient}
   \tag{2}
\end{equation}
As shown in Figure 5d, only the values of C for QF-MP(r\textsubscript{q}=1.2μm) is larger than 0.2, that means the strong coupling between SPhPs of quartz micropillars and the molecular vibrations of CO\textsubscript{2} molecules may not be achieved in QF-MP with other r\textsubscript{q} value.
\begin{figure}
    \centering
    \includegraphics[width=1\linewidth]{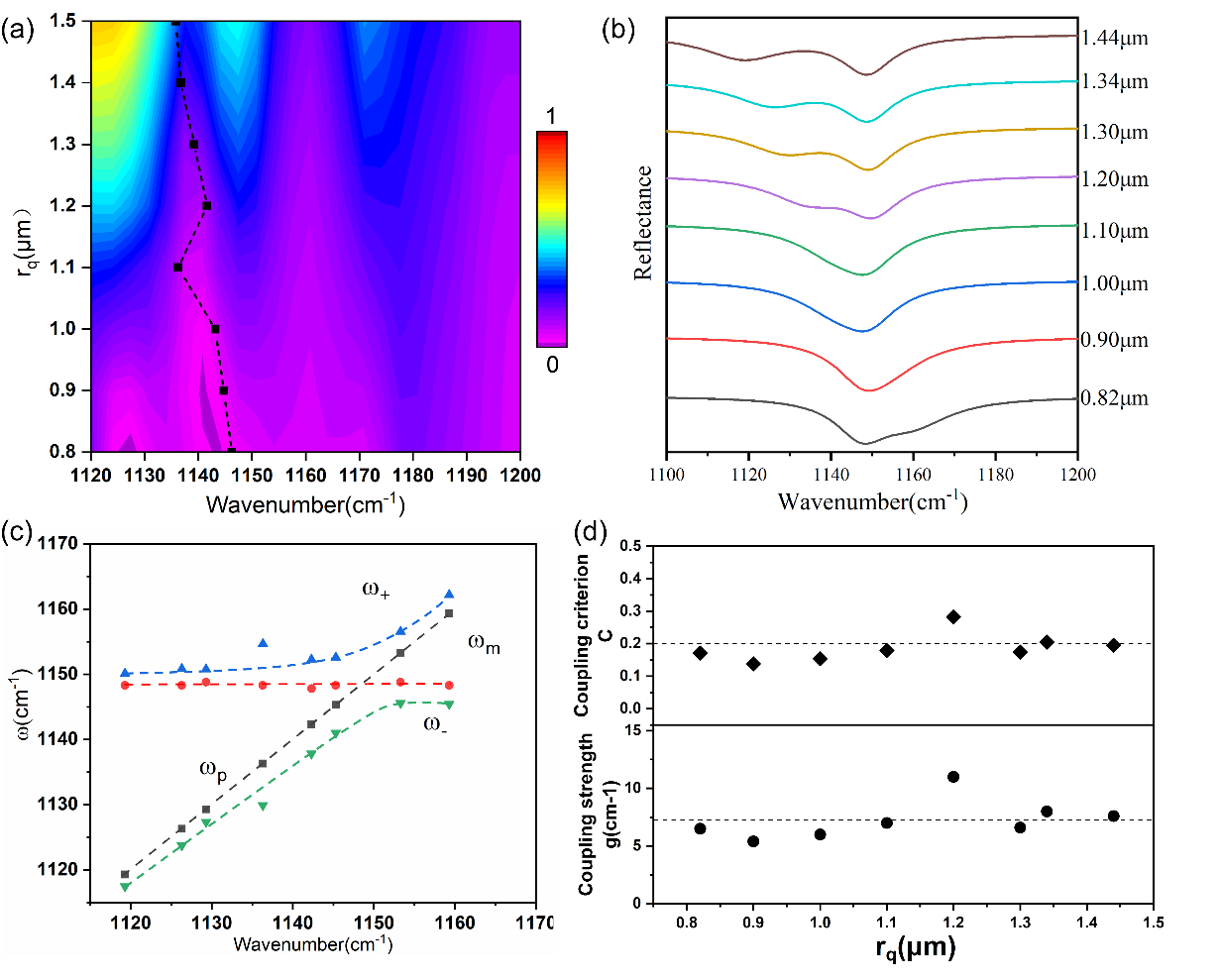}
Figure 4. (a) The “pinning” effect of the mid-infrared SPhP resonances. The color map represents the simulated reflectance of micro-quartz pillar arrays. The black dots represent the measured resonance positions in Fig 4f; (b) Experimental reflectance of QF-MP with N-CAF catalyst under strong coupling between mid-infrared SPhP and the CO\textsubscript{2} vibrational mode; (c) Eigen hybrid-mode frequencies ω\textsubscript{±} of the coupling system as a function of the bare SPhP frequency. The ω\textsubscript{p} and ω\textsubscript{m} used in the coupled oscillator fitting are also plotted. d) The coupling strength g and the criterion C as a function of r\textsubscript{q}.
\end{figure}[H]

\subsection{The performances for photocatalytic CO\textsubscript{2} reduction}

Cu\textsubscript{2}O has a narrow band gap (∼1.9 eV), low conductivity, and poor stability due to the facile redox of Cu\textsuperscript{+}, which restricts its photocatalytic performance\textsuperscript{23}. To overcome these drawbacks, we constructed a composite catalyst (denoted N-CAF) by nitrogen doping and integrating Ag and FeO\textsubscript{x}. N doping narrows the band gap and reduces resistivity by up to two orders of magnitude, while Ag nanoparticles provide surface plasmon resonance to enhance light absorption and carrier separation\textsuperscript{24-26}. The p–n heterojunction between Cu\textsubscript{2}O and Fe\textsubscript{2}O\textsubscript{3} further improves charge separation and oxidation ability.

XRD confirmed the coexistence of cubic Cu\textsubscript{2}O, Fe\textsubscript{2}O$_3$/Fe$_3$O$_4$, and metallic Ag phases (Fig S8). SEM/TEM images revealed well-defined cubic Cu$_2$O particles (50–100 nm) interfaced with FeO$_x$ spheres, with Ag nanoparticles uniformly anchored on Cu$_2$O surfaces (Fig 5). Lattice fringe analysis identified Cu$_2$O (100), Ag (111), and Fe$_2$O$_3$ planes, indicating strong interfacial contact favorable for carrier transport. Elemental mapping further confirmed the homogeneous distribution of Ag, Fe, Cu, O, and N.

XPS spectra verified the coexistence of Cu\textsuperscript{+}, Fe\textsuperscript{3+}/Fe\textsuperscript{2+}, and metallic Ag, while N 1s peaks indicated multiple N bonding environments (Cu–O–N, N–O, and Cu–N–Cu), beneficial for electron transfer and CO\textsubscript{2} adsorption (Fig S8). Collectively, these results demonstrate the successful fabrication of N-CAF with a well-defined Cu\textsubscript{2}O/FeO\textsubscript{x} heterojunction, uniformly doped N, and plasmonic Ag nanoparticles, providing a favorable architecture for photocatalysis.
\begin{figure}[H]
    \centering
    \includegraphics[width=1\linewidth]{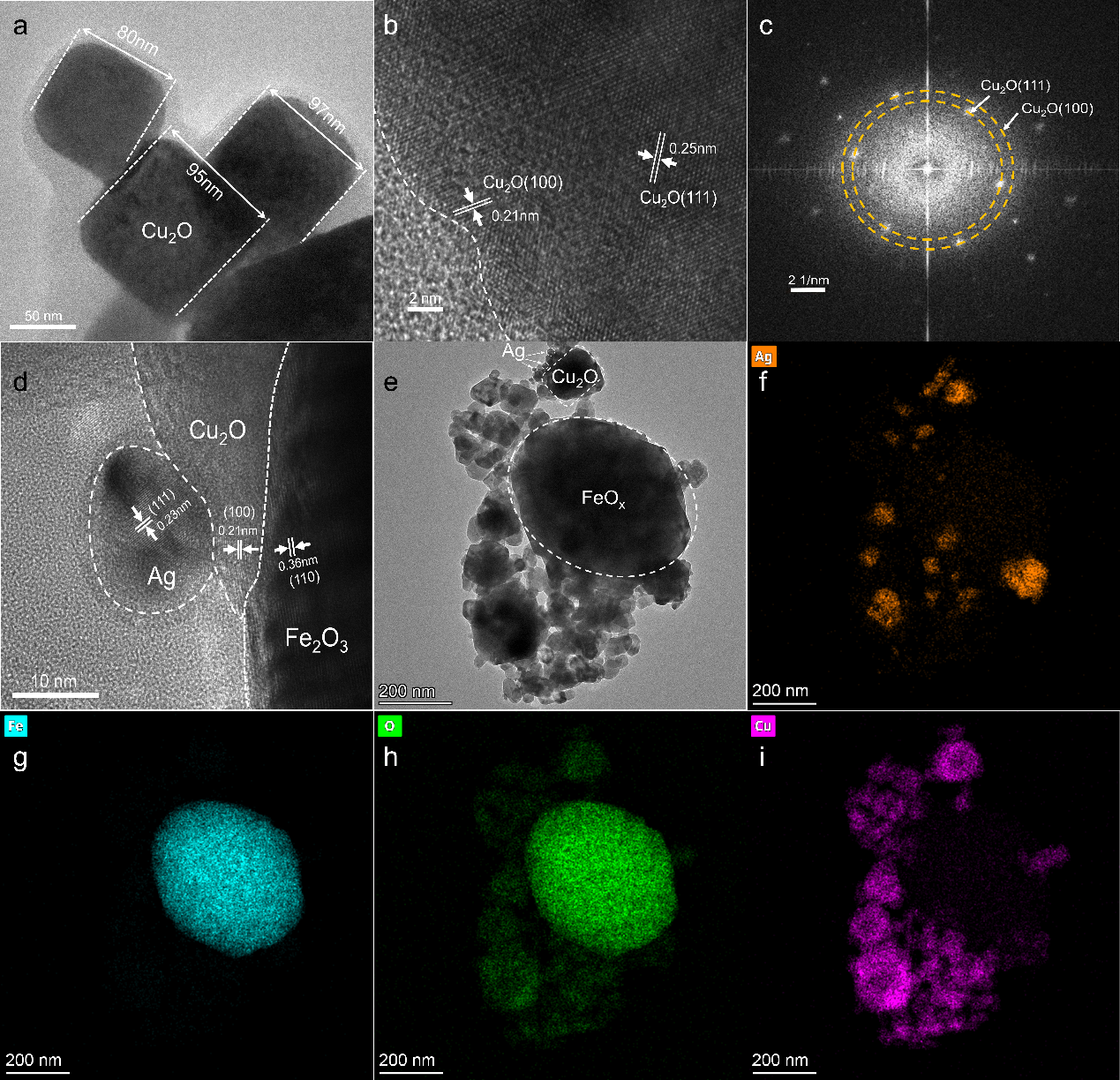}
Figure 5. TEM images and SAED patterns of a-c) Cu\textsubscript{2}O, d), and e) N-CAF. EDX mapping images of f) Ag, g) Fe, h) O, and i) Cu elements of N-CAF as shown in e).
\end{figure}
We conducted the CO\textsubscript{2}RR experiments using a custom-designed photocatalytic in-situ detection system, as shown in Fig S6. All experiments were carried out with glass reactor using a Φ40×5 mm CaF\textsubscript{2} window in a pure water environment.

Firstly we compared the photocatalytic yield and selectivity of Cu\textsubscript{2}O, N-Cu\textsubscript{2}O, FeO\textsubscript{x}, CAF, and N-CAF catalysts loaded onto 1×1 cm quartz substrates under Xeon lamp irradiation, as presented in Fig S6. After N doping, the CH\textsubscript{4} yield of N-Cu\textsubscript{2}O increased, while the CO yield decreased, with CO selectivity dropping from 80\% to 48\%. According to the calculations in Fig S10, N doping enhances the CO\textsubscript{2} adsorption ability of the catalyst, as confirmed by BET tests (Fig S5), leading to an increase in the sum yield. However, N doping also increases the adsorption energy of *CO, making it harder for CO to desorb from the surface, favoring its subsequent reaction with surface *H to produce CH\textsubscript{4}. FeO\textsubscript{x} exhibited better catalytic activity and lower CO selectivity (26\%). 

The CO yield of CAF (108.4μmol/h/g) was significantly higher than that of the first three catalysts, while its CH\textsubscript{4} yield was lower than that of FeO\textsubscript{x}, indicating that the Z-scheme heterojunction formed between Cu\textsubscript{2}O and FeO\textsubscript{x} allows Cu\textsubscript{2}O to consume part of the electrons from FeO\textsubscript{x} as the reduction site, suppressing the reduction ability of FeO\textsubscript{x} (Fig 6a). The overall yield of N-CAF was higher than that of CAF, but its CO selectivity decreased from 77.5\% to 67.2\%. This trend is consistent with the changes observed in N-doped Cu\textsubscript{2}O, although the yield of CO production was relatively decrease. This may be attributed to the plasmonic effect of Ag nanoparticles, which increases the surrounding temperature, facilitating CO desorption from the surface. 

As shown in Fig 6a, we performed control experiments on the CO\textsubscript{2}RR yield of N-CAF loaded onto QF-MP (rq=1.2 μm) under different lighting conditions. No CO or CH\textsubscript{4} products were detected under dark conditions or with infrared lamp irradiation alone. Under simultaneous irradiation by the infrared lamp and Xeon lamp, the CO yield increased from 113.6μmol/h/g to 167.7μmol/h/g. Combined with prior calculations and characterizations, this indicates that the formation of VSC between SPhP and CO\textsubscript{2} molecules can enhance the CO\textsubscript{2}RR reaction rate. The CH\textsubscript{4} yield improvement was less pronounced, possibly because the increased temperature made CO more likely to desorb, and the C=O bond in CO may have weak coupling with SPhP, making it more effective in absorbing energy transferred by phonons. 

We further tested the CO2RR performance of N-CAF on different substrates under Xeon and infrared lamp irradiation. As shown in Fig 6c, other QF-MP substrates, except QF-MP (rq=1.2 μm), did not show significantly higher yields compared to ordinary quartz substrates. This suggests that the vibrational coupling between SPhP and CO\textsubscript{2} molecules must reach a certain intensity to activate CO\textsubscript{2} molecules and enhance the CO\textsubscript{2}RR reaction rate. 

To rule out the thermal effects caused by infrared lamp irradiation, we measured the surface temperature of N-CAF loaded on QF-MP (r\textsubscript{q}=1.2 μm) under different conditions: Xeon lamp irradiation and both Xeon and infrared lamp irradiation. The results showed that the equilibrium temperature of the latter was only 1.5℃ higher than one of the former. 
\begin{figure}[H]
    \centering
    \includegraphics[width=1\linewidth]{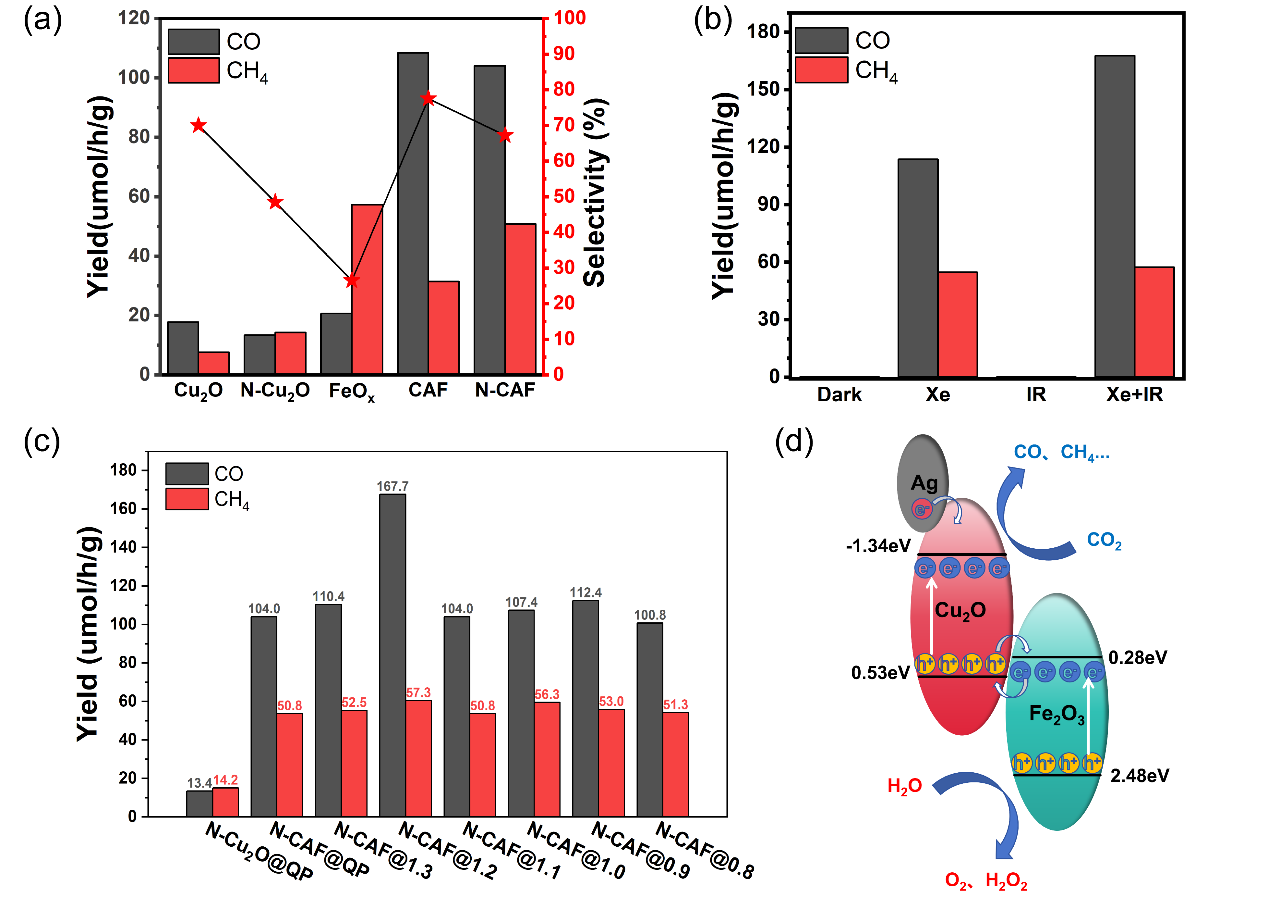}
Figure 6. (a) product yields and the selectivity of CO with Cu\textsubscript{2}O, N-Cu\textsubscript{2}O, FeO\textsubscript{x}, CAF and N-CAF loaded on quartz film under Xeon lamp, b) the CO\textsubscript{2}RR yield of N-CAF@QF-MP(rq=1.2μm) under differet condition: dark, Xeon lamp, IR lamp and both lamps, c) the CO\textsubscript{2}RR yield N-CAF@QF-MP(rq=0.8\~1.3um), d) a proposed mechanism for photocatalytic reduction of CO\textsubscript{2} catalyzed by N-CAF.
\end{figure}

\subsection{Conclusion}

To address the challenge of activating and dissociating CO₂ molecules adsorbed on catalyst surfaces during photocatalytic CO₂ reduction, this study integrates theoretical calculations with experimental design. Density functional theory (DFT) calculations reveal that CO₂ exhibits weak adsorption on the stable Cu₂O (100) surface. To overcome this limitation, nitrogen doping was introduced to generate active adsorption and reaction sites, thereby enhancing CO₂ adsorption energy and lowering the dissociation energy barrier. Computational analysis of the adsorbed “armchair” CO₂ configuration further identified its asymmetric stretching vibration at \~1136 cm⁻¹ as the key vibrational mode associated with surface dissociation.

To amplify this vibrational activation, micro-pillar arrays were fabricated on α-quartz substrates to induce surface phonon-polariton (SPhP) effects. By tailoring the pillar dimensions, resonance frequencies were tuned within the 1072–1215 cm\textsuperscript{-1} range, matching the CO\textsubscript{2} vibrational mode. Moreover, an N-doped Cu\textsubscript{2}O/Ag/Fe\textsubscript{2}O\textsubscript{3} (N-CAF) catalyst was constructed, which delivered enhanced photocatalytic activity with CO and CH\textsubscript{4} yields of 113.6 μmol g\textsuperscript{-1} h\textsuperscript{-1} and 54.6 μmol g\textsuperscript{-1} h\textsuperscript{-1}, respectively. When the N-CAF catalyst was supported on quartz micro-pillars (QF) and subjected to combined Xe and infrared irradiation in pure water, the yields further increased to 167.7 μmol g\textsuperscript{-1} h\textsuperscript{-1} for CO and 57.3 μmol g\textsuperscript{-1} h\textsuperscript{-1} for CH\textsubscript{4}. Notably, vibrational strong coupling (VSC) between molecular vibrations and SPhP modes was observed, resulting in a 47.6\% improvement in CO yield compared to systems without VSC.

This work demonstrates microscopic regulation of CO₂ activation and dissociation in photocatalysis via VSC. By harnessing light–vibration–molecule coupling, the reaction energy barrier is effectively reduced, thereby improving energy utilization and conversion efficiency. The strategy provides a novel pathway to enhance CO₂ reduction yields and advance photo–thermal synergistic catalysis.

\section{Methods}

\textbf{
\subsection{Fabrication of Substrate}
}
Quartz substrates (<0001>, Hefei Kejing) were first plasma-treated and coated with a 100 nm Cr layer by magnetron sputtering. Micro-pillar patterns were then defined using standard photolithography with AZ6112 photoresist, followed by ion beam etching (IBE) of the Cr mask and subsequent SiO\textsubscript{2} etching (2.0 μm) via ICP-RIE using CF\textsubscript{4}/Ar plasma. After resist and Cr mask removal through organic cleaning and wet etching, the patterned quartz substrates were diced into individual cells using a DISCO cutting system. (The details is shown in S1 section)

\textbf{
\subsection{Optical Properties Simulation}
}
The ’Electromagnetic Waves, Frequency Domain’ module in the COMSOL\textregistered was used to calculate the optical properties of plasmonic materials. We built a period models to simulate structure of the quartz film with micro-pillar array. Due to the surface scattering, grain boundary effects, and the gallium ion contamination in thin films, the damping constant of the quartz film is set to be five times as that of bulk quartz in the simulation.

\subsection{Catalyst Synthesis
}
Synthesis of FeO\textsubscript{x} :The FeO\textsubscript{x} photocatalysts is successfully synthesized by four method: i) add 50 mL of 2 M FeCl\textsubscript{3} solution dropwise into 50 mL of 5.4 M NaOH; ii) stir at 75°C for 5 minutes at 500 rpm, then wash four times with deionized water and ethanol;iii) dry the product in an oven at 100°C for 48 hours; iv) anneal it at 500°C under an Ar atmosphere for 4 hours.

Synthesis of Cu\textsubscript{2}O: The Cu\textsubscript{2}O was synthesized via the reduction reaction of CuSO\textsubscript{4} as follow step. First, CuSO\textsubscript{4}·5H\textsubscript{2}O (1.5 mmol) was dissolved in 100 mL of deionized water. Then, Na\textsubscript{3}C\textsubscript{6}H\textsubscript{5}O\textsubscript{7}·2H\textsubscript{2}O (0.51 mmol) and NaOH (25 mmol) were sequentially added to the CuSO\textsubscript{4}·5H\textsubscript{2}O solution, followed by stirring for 5 minutes to obtain a homogeneous solution. Separately, ascorbic acid (1.5 mmol) was dissolved in 50 mL of deionized water and mixed thoroughly. The ascorbic acid solution was then dropwise added to the above mixture. After vigorous stirring at room temperature for 40 minutes, the final product was collected by centrifugation and washed several times with ethanol and deionized water. Finally, the product was dried in an oven at 80°C for 12 hours.

Synthesis of N-Cu\textsubscript{2}O: For the synthesis of N-Cu\textsubscript{2}O and N, the prepared Cu\textsubscript{2}O was annealed under NH\textsubscript{3} (99.99\%) atmosphere. The sample was heated at a rate of 5°C/min to 280°C and held at this temperature for 30 minutes.

Synthesis of Cu\textsubscript{2}O/Ag/FeO\textsubscript{x} : Add the FeO\textsubscript{x} prepared in S6.1 to 200 mL DI water and sonicate for 1 hour. Then, CuSO\textsubscript{4}·5H\textsubscript{2}O (1.5 mmol) was dissolved in the solution. The next step is add Na\textsubscript{3}C\textsubscript{6}H\textsubscript{5}O\textsubscript{7}·2H\textsubscript{2}O (0.51 mmol) and NaOH (25 mmol) added to the solution sequentially, followed by stirring for 5 minutes to obtain a homogeneous solution. Separately, ascorbic acid (1.5 mmol) was dissolved in 50 mL of deionized water and mixed thoroughly. The ascorbic acid solution was then dropwise added to the above mixture. After vigorous stirring at room temperature for 40 minutes, the final product was collected by centrifugation and washed several times with ethanol and deionized water. Then, add the separated powder back into 200 mL DI water and sonicate for 30 minutes.

Add 0.1 M NaNO\textsubscript{3} and 0.1 M AgNO\textsubscript{3} into the solution. Under dark conditions, stir at 700 r/min for 15 minutes, followed by irradiation with a UV lamp (18W mercury lamp, 253 nm) for 20 minutes. After washing and drying, place the sample in a tube furnace and anneal in an Ar atmosphere at 400°C for 4 hours.

\textbf{
\subsection{Catalyst Characterizations}
}
The crystallographic phase of these as-prepared powder samples was determined investigated by an in-situ temperature-dependent and ex-situ X-ray diffractometer (PANalytical X’Pert3 Powder) at different temperatures. The rate of rising temperature is 10 °C/min and maintain the test temperature 5 min. The XRD graphs were described over the scanning range of 5°-80° using Cu-Ka radiation (λ = 0.154178 nm) at 40 kV and 40 mA. The morphologies of the materials and EDS mapping were determined by a field emission scanning electron microscope (FESEM) by using an accelerating voltage of 5 kV. Transmission electron microscopic (TEM) images were obtained at 200 kV by a transmission electron microscope. The X-ray photoelectron spectroscopy (XPS) was performed using a spectrometer (Escalab 250xi, Thermo Scientific). The specific surface area and pore size distribution of materials was performed by a physical absorption analyzer. 

 \textbf{
\subsection{In-situ DRIFTS Measurements}
}
In-situ diffuse reflectance Fourier transform infrared spectroscopy (DRIFTS) was conducted using a Bruker IFS 66v spectrometer equipped with a Harrick diffuse reflectance accessory at the Infrared Spectroscopy and Microspectroscopy Endstation (BL01B) of NSRL. Spectra were collected at a resolution of 2 cm\textsuperscript{-1} by averaging 256 scans. The catalysts were loaded into an in-situ infrared chamber with ZnSe windows, specifically designed for highly scattering powder samples. Prior to measurements, the chamber was purged with Ar for 30 min, and the resulting spectrum was recorded as the background. During in-situ experiments, a CO\textsubscript{2}/H\textsubscript{2}O mixture was continuously introduced into the chamber.

 \textbf{
\section{Funding:}
}
This work was financially supported by the Basic Science Center Program for Ordered Energy Conversion of the National Natural Science Foundation of China (No. 52488201)

\textbf{
\section{Conflicts of Interest:}
}
There are no competing financial interests to be declared.

\textbf{
\section{Supporting Information}
}
\textbf{
\subsection{S1 Fabrication of Quartz Films}
}

The specific steps are as follows:

a) Quartz sheet (Hefei Kejing, <0001>)
b) Substrate treatment plasma treatment M4L plasma adhesive remover 400W5min
c) coating Cr100nm equipment: FHR magnetron sputtering table, Cr: power 300w, rate 3.2A/s, coating 100nm
d) mask plates, 5 inches
e) lithography Mask1 AZ6112 photoresist, uniform glue: 600r * 5s, 4000r * 30s, front baking: 100 degrees for 2min, exposure: MA6, exposure for 2s, development: 2.38\% TMAH developer, development for 30s, firm film: 100 degrees for 3min
f) IBE etching CrPt IBE ion beam etching machine: working vacuum: 2E-2Pa, ion energy: 350eV, ion beam current: 80mA, etching rate: Cr4nm/min
g) 7 NLD etching SIO\textsubscript{2} CF\textsubscript{4}:40sccm, Ar: 10sccm, ICP1500W, RIE150W, etching 2.0 um
h) Gel removal cleaning, organic cleaning, acetone+isopropanol ultrasonic cleaning for 5 minutes
i) Wet corrosion of Cr by removing Cr. Cr corrosion solution corrodes the remaining Cr
j) Cut DISCO Cut according to Cell Cut

 \textbf{
\subsection{S2 DFT computation details}
}

We carried out the first-principle simulations in the framework of DFT using the generalized gradient approximation (GGA) of Perdew–Burke–Ernzerhof (PBE)\textsuperscript{27}. The VASP (Vienna ab initio simulation package Version 6.3.0) package was employed with the projected augmented-wave method\textsuperscript{28,29}. The kinetic cutoff energy for the plane-wave basis is 520 eV. The Brillouin zone integration was performed on a $\Gamma$-centered with K-Spacing value equal to 0.04. In the geometric structure optimization calculations, all the atoms were fully relaxed until the force on each atom was less than 0.02 eV /Å(0.05 eV /Å in CI-NEB). To analyze the properties of different material surfaces, we used the periodic slab models with a vacuum layer of 15 Å. And the bottom layers of atoms on the bottom were fixed. Transition state searches were calculated using the climbing image nudged elastic band\textsuperscript{30,31}.

\textbf{
\subsection{S3 The permittivity of α-quartz}
}

The permittivity of α-quartz in the infrared spectral regime is described by the Drude model, as follows\textsuperscript{32}:
\begin{equation}
    \varepsilon_{\text{quartz}} = \varepsilon_{\infty} \left( 1 + \frac{\omega_{\text{LO}}^2 - \omega_{\text{TO}}^2}{\omega_{\text{TO}}^2 - i \omega \gamma - \omega^2} + \frac{\omega_1^2 - \omega_2^2}{\omega_2^2 - i \omega \gamma - \omega^2} \right)
    \label{eq:quartz_dielectric_function}
\tag{S1}
\end{equation}
where $\varepsilon$\textsubscript{$\infty$} = 2.356, ω\textsubscript{LO}=1215 cm\textsuperscript{-1}, ω\textsubscript{TO}= 1072 cm\textsuperscript{-1}, γ = 7.6 cm\textsuperscript{-1}, ω\textsubscript{1} = 1165 cm\textsuperscript{-1}, ω\textsubscript{2} = 1163 cm\textsuperscript{-1}, γ\textsubscript{1} = 7 cm\textsuperscript{-1}. The permittivity is plotted in Figure S1.
\begin{figure}
    \centering
    \includegraphics[width=1\linewidth]{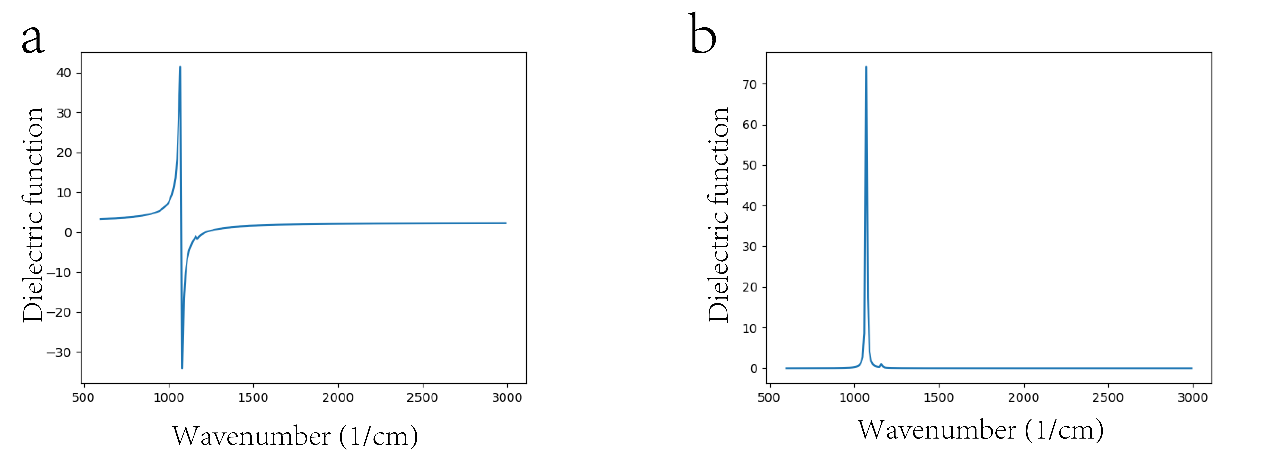}
\textbf{Figure S1}. The permittivity of α-quartz, (a) real ($\varepsilon$\textsubscript{1}) part of the permittivity of α-quartz, (b) imaginary ($\varepsilon$\textsubscript{2}) part of the permittivity of α-quartz.
\end{figure}

 \textbf{
\subsection{S4 The harmonic oscillator modelling}
}

In order to analyze the infrared reflectance spectra shown in Figure 5a of the main text , we described the coupling of the SPhP resonances and the molecular vibrations via a classical model of coupled harmonic oscillators. The equations of motion for the two coupled harmonic oscillators are:
\[{\ddot{x}}_{p}(t) + \gamma_{p}{\dot{x}}_{p}(t) + \omega_{p}^{2}x_{p}(t) - g\left( \omega_{p} + \omega_{m} \right)x_{m}(t) = F_{p}(t)\]

\[\begin{array}{r}
{\ddot{x}}_{m}(t) + \gamma_{m}{\dot{x}}_{m}(t) + \omega_{m}^{2}x_{m}(t) - g\left( \omega_{p} + \omega_{m} \right)x_{p}(t) = F_{m}(t)\
\tag{S2}
\end{array}\]
Where  \(x_{p}\),\(\omega_{p}\),\(\gamma_{p}\),\(F_{p}\) represent the displacement, frequency, damping and the effective force driving oscillator motion of the plasmonic resonance, respectively. The corresponding notation is also appropriate for the molecule of CO\textsubscript{2}. And g is the coupling strength.

The reflectance spectra was fitted by: 
\[\begin{array}{r}
R = 1 - \sigma\
\tag{S3}
\end{array}\]
where \(\sigma \propto \left\langle F_{p}(t){\dot{x}}_{p}(t) + F_{m}(t){\dot{x}}_{m}(t) \right\rangle\). In the fitting, \(\omega_{m}\) (\(\omega_{m}\) = 1036 ± 0.5 cm\textsuperscript{-1}) and \(\gamma_{m}\) (\(\gamma_{m}\) = 18.0 ± 0.1 cm\textsuperscript{-1}) were allowed to fluctuate within a small range. We considered ω\textsubscript{p} and γ\textsubscript{p} as free parameters, because the SPhP resonance has a shift in peak positions and a modify in linewidths due to the change of the dielectric environment once the molecules were coated on the micro-pillars of quartz film. The fitted parameter results are shown in Table S3.

 \textbf{
\subsection{S5 Tests on photocatalytic CO\textsubscript{2} reduction with H\textsubscript{2}O}
}

1.5 mg of the photocatalyst with glass fiber filter and 2 mL deionized water was placed in a 50 mL quartz reactor. Besides, a 300 W Xenon lamp and a 650mW Infrared light source were used as reactive light source. Firstly, high purity (99.999 \%) CO\textsubscript{2} with a steady flow (6 scccm) was continuously bubbled in the closed reactor to ensure no O\textsubscript{2} and N\textsubscript{2} in the reactor and the CO\textsubscript{2}/H\textsubscript{2}O adsorption-desorption equilibrium on the photocatalyst for at least 2 h, and then start the light reaction. After irradiation for 1 h, 1 mL mixed gas of the photo-reactor was injected into a gas chromatograph (GC-9720, containing both TCD and FID detectors) every 30 minutes to obtain CO and CH\textsubscript{4} concentrations. Meanwhile, control experiments were also done in Ar atmosphere, no light or no catalyst. At least 13 samples are collected in each experiment, and the results of the first 4 samples are not considered valid data.

 \textbf{ 
\subsection{S6 Calculations of AQE:}
}

Primary product: CH\textsubscript{4}
Product yield = 57.3 μmol$\cdot$g\textsuperscript{-1}$\cdot$h\textsuperscript{-1}
Apparent Xenon Light input (H\textsubscript{x}) = 1000 W$\cdot$m\textsuperscript{-2}
Area of reactor under irradiation of Xenon light (A) = 0.0001 m\textsuperscript{2}
Band gap (E\textsubscript{g}) = 1.89 eV
Apparent Infrared light input (H\textsubscript{i}) = 35W$\cdot$m\textsuperscript{-2}

The calculations to find the electrons, participated in photocatalytic reaction, are as follow
\[\text{number of reacted electrons = }\left\lbrack \text{Production rate of }{CH}_{4} \right\rbrack\text{ ×}\left\lbrack \begin{array}{r}
\text{electrons required per } \\
\text{mole of }{CH}_{4}\text{ }formation
\end{array} \right\rbrack\text{× }\text{N}_{\text{A}}\]
As we know from balanced chemical equation, CO\textsubscript{2} + 8H\textsuperscript{+} + 8e\textsuperscript{‑} \(\rightarrow\) CH\textsubscript{4} + 2H\textsubscript{2}O, 8 electrons are consumed per mole of CH\textsubscript{4} formed, therefore
\[\text{number of reacted electrons =}\left\lbrack \text{5.73×}\text{10}^{\text{-5}}\text{ mol$\cdot$}g^{- 1}\text{$\cdot$}h^{- 1} \right\rbrack\text{× }\left\lbrack \text{8} \right\rbrack\text{ × 6.022×}\text{10}^{\text{23}}\text{  }{mol}^{- 1}\]
\[\text{number of reacted electrons = 2.76×}\text{10}^{\text{20}}\ g^{- 1}\text{$\cdot$}h^{- 1}\]
\[\text{Basis for calculations = 3 mg}\]
\[\text{number of reacted electrons = 8.28×}\text{10}^{\text{17}}\ h^{- 1}\]
The number of photons incident upon photocatalyst is calculated by
\[\text{number of incident photons = }\frac{\text{light absorbed by the photocatalyst}}{\text{average energy of the photon }}\text{ × t}\]
Light absorbed by the photocatalyst under the Xeon lamp and average energy of photon is calculated as
\begin{quote}
\[\text{light absorbed by the photocatalyst = }H_{x}\text{ × A = 1000 }{W\text{$\cdot$}m}^{- 2}\text{ × 0.0001 }\text{m}^{\text{2}}\text{ = 0.1 W}\ \]
\[\text{average energy of the photon = }\frac{\text{hc}}{\text{λ}}\]
\end{quote}
Where, h is Planck's constant (6.626\(\text{×}\)10\textsuperscript{-34} J$\cdot$s), c is speed of light (3\(\text{×}\)10\textsuperscript{8}  m$\cdot$s\textsuperscript{-1}) and λ is the average wavelength for the photocatalyst\textquotesingle s absorption range.

For finding λ we calculated λ\textsubscript{max} from the band gap by the formula:
\begin{quote}
\[\text{λ}_{\text{max}}\text{ = }\frac{\text{hc}}{\text{E}_{\text{g}}}\text{ = }\frac{\left( \text{6.626×}\text{10}^{\text{-34}}\text{ J$\cdot$s} \right)\text{ × }\left( \text{3×}\text{10}^{\text{8}}\text{ m$\cdot$}s^{- 1} \right)}{\text{1.90 eV}}\text{ ×}\frac{\text{1 eV}}{\text{1.6×}\text{10}^{\text{-19}}\text{ J}}\text{ ×}\text{ 10}^{\text{9}}\text{nm$\cdot$}m^{- 1}\text{ = 653 nm}\]
\end{quote}
Therefore, the average wavelength would be
\[\text{λ = }\frac{\text{λ}_{\text{min}}\text{ + }\text{λ}_{\text{max}}}{\text{2}}\text{ = }\frac{\text{250 +653}}{\text{2}}\text{ = 451.5 nm}\]
The average photon energy is then computed to be
\begin{quote}
\[\text{average photon energy = }\frac{\left( \text{6.626×}\text{10}^{\text{-34}} \right)\text{ × }\left( \text{3×}\text{10}^{\text{8}} \right)}{\text{451.5×}\text{10}^{\text{-9}}\text{ m}}\text{ = 4.40×}\text{10}^{\text{-19}}\text{ J}\]
\end{quote}
\[\text{number of incident photons =}\ \frac{\text{0.1 W}}{\text{4.40×}\text{10}^{\text{-19}}\text{ J}}\text{ × 3600 s$\cdot$}h^{- 1}\]
\[\text{number of incident photons = 8.18×}\text{10}^{\text{21}}\text{ }h^{- 1}\]
AQE computation:
\[\text{AQE }\left( \text{\%} \right)\text{ = }\frac{\text{number of reacted electrons}}{\text{number of incident photons}}\text{ × 100\%}\]
\[\text{AQE}\ \left( \text{\%} \right)\text{ =}\ \frac{\text{8.28×}\text{10}^{\text{17}}}{\text{8.18×}\text{10}^{\text{21}}}\text{ × 100\% = 0.01\%}\]
\begin{quote}
Primary product: CO
Product yield = 167.7 μmol\(\text{$\cdot$}\)g\textsuperscript{-1}\(\text{$\cdot$}\)h\textsuperscript{-1}
Apparent Light input (H) = 1000 W\(\text{$\cdot$}\)m\textsuperscript{-2}
Area of reactor under irradiation (A) = 0.0001 m\textsuperscript{2}
Band gap (E\textsubscript{g}) = 1.89 eV
\end{quote}
The calculations to find the electrons, participated in photocatalytic reaction, are as follow
\[\text{number of reacted electrons = }\left\lbrack \text{Production rate of }{CH}_{4} \right\rbrack\text{ ×}\left\lbrack \begin{array}{r}
\text{electrons required per } \\
\text{mole of }{CH}_{4}\text{ }formation
\end{array} \right\rbrack\text{× }\text{N}_{\text{A}}\]
As we know from balanced chemical equation, CO\textsubscript{2} + 2H\textsuperscript{+} + 2e\textsuperscript{‑} \(\rightarrow\) CO + H\textsubscript{2}O, 2 electrons are consumed per mole of CO formed, therefore
\[\text{number of reacted electrons =}\left\lbrack \text{167.7×}\text{10}^{\text{-5}}\text{ mol$\cdot$}g^{- 1}\text{$\cdot$}h^{- 1} \right\rbrack\text{× }\left\lbrack \text{2} \right\rbrack\text{ × 6.022×}\text{10}^{\text{23}}\text{  }{mol}^{- 1}\]
\[\text{number of reacted electrons = 2.02×}\text{10}^{\text{21}}\ g^{- 1}\text{$\cdot$}h^{- 1}\]
\[\text{Basis for calculations = 3 mg}\]
\[\text{number of reacted electrons = 6.06×}\text{10}^{\text{18}}\ h^{- 1}\]
AQE computation:
\[\text{AQE }\left( \text{\%} \right)\text{ = }\frac{\text{number of reacted electrons}}{\text{number of incident photons}}\text{ × 100\%}\]
\[\text{AQE}\ \left( \text{\%} \right)\text{ =}\ \frac{\text{6.06×}\text{10}^{\text{18}}}{\text{8.18×}\text{10}^{\text{21}}}\text{ × 100\% = 0.074\%}\]
\textbf{
\subsection{S7 Figure}
}
\begin{figure}[H]
    \centering
    \includegraphics[width=0.5\linewidth]{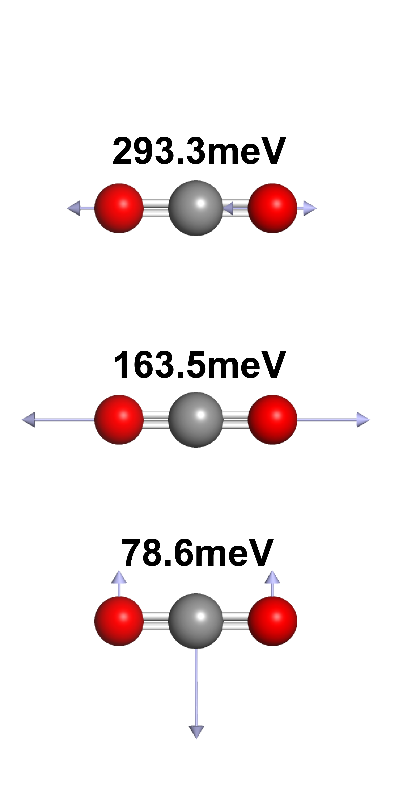}
\caption*{Figure S2. The main vibrational modes of gaseous CO\textsubscript{2} molecules}
\end{figure}

\begin{figure}[H]
    \centering
    \includegraphics[width=0.75\linewidth]{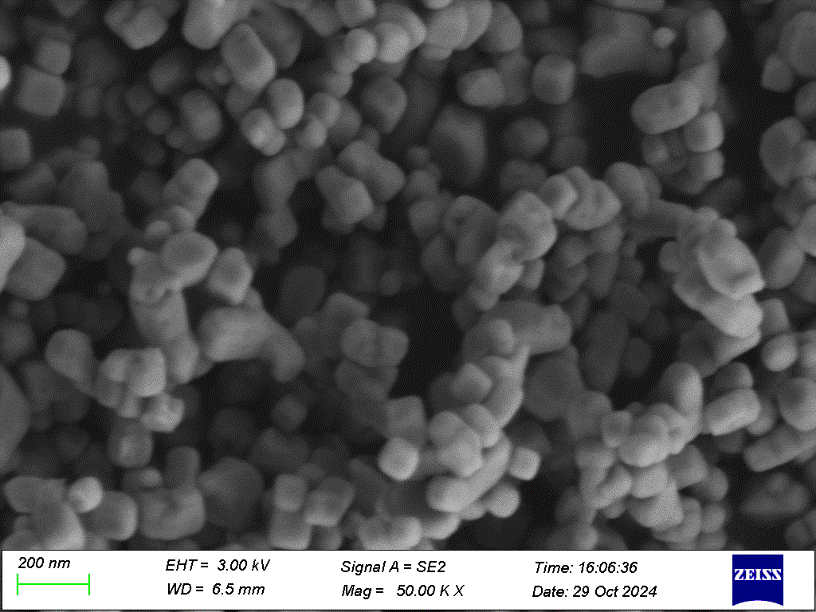}
\caption*{Figure S3. SEM of Cu\textsubscript{2}O powder}
\end{figure}

\begin{figure}[H]
    \centering
    \includegraphics[width=1\linewidth]{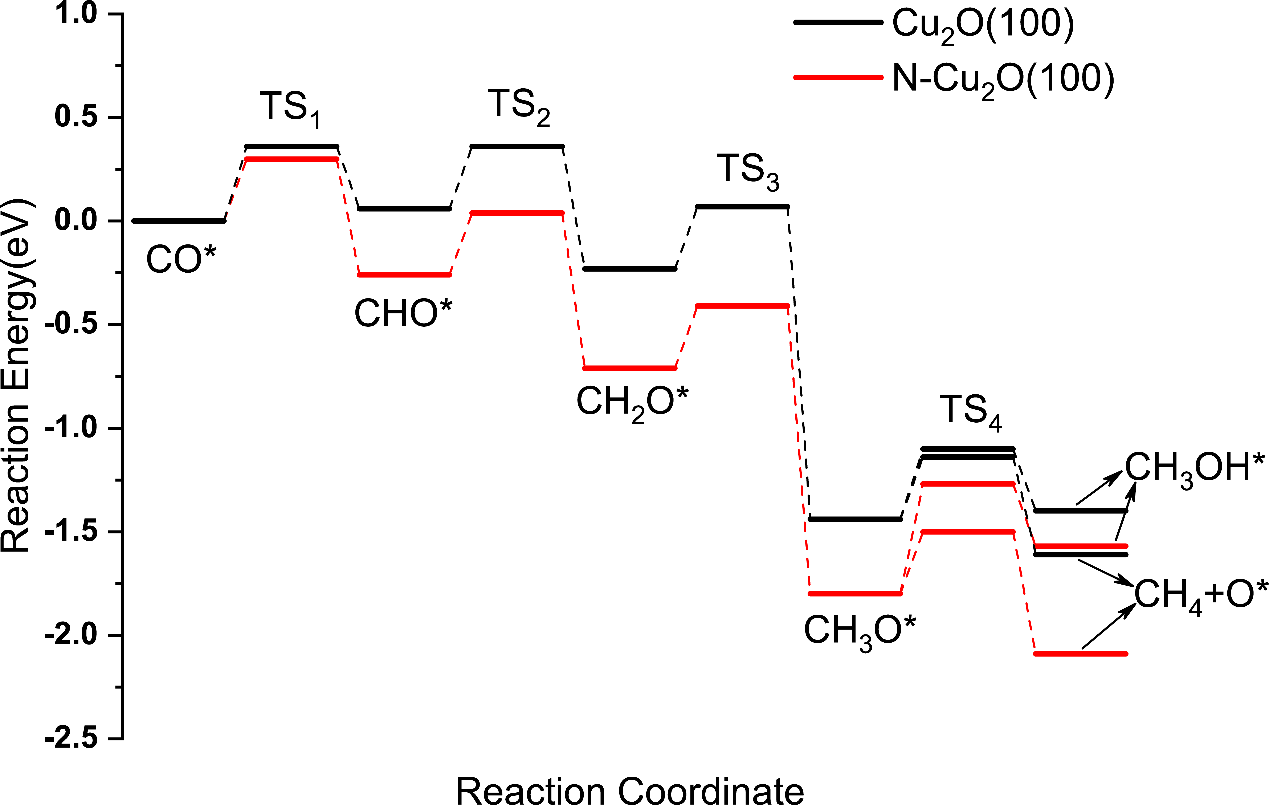}
\caption*{Figure S4. CO\textsubscript{2}RR Gibbs energy diagram computed for Cu\textsubscript{2}O and N-Cu\textsubscript{2}O}
\end{figure}

\begin{figure}[H]
    \centering
    \includegraphics[width=0.75\linewidth]{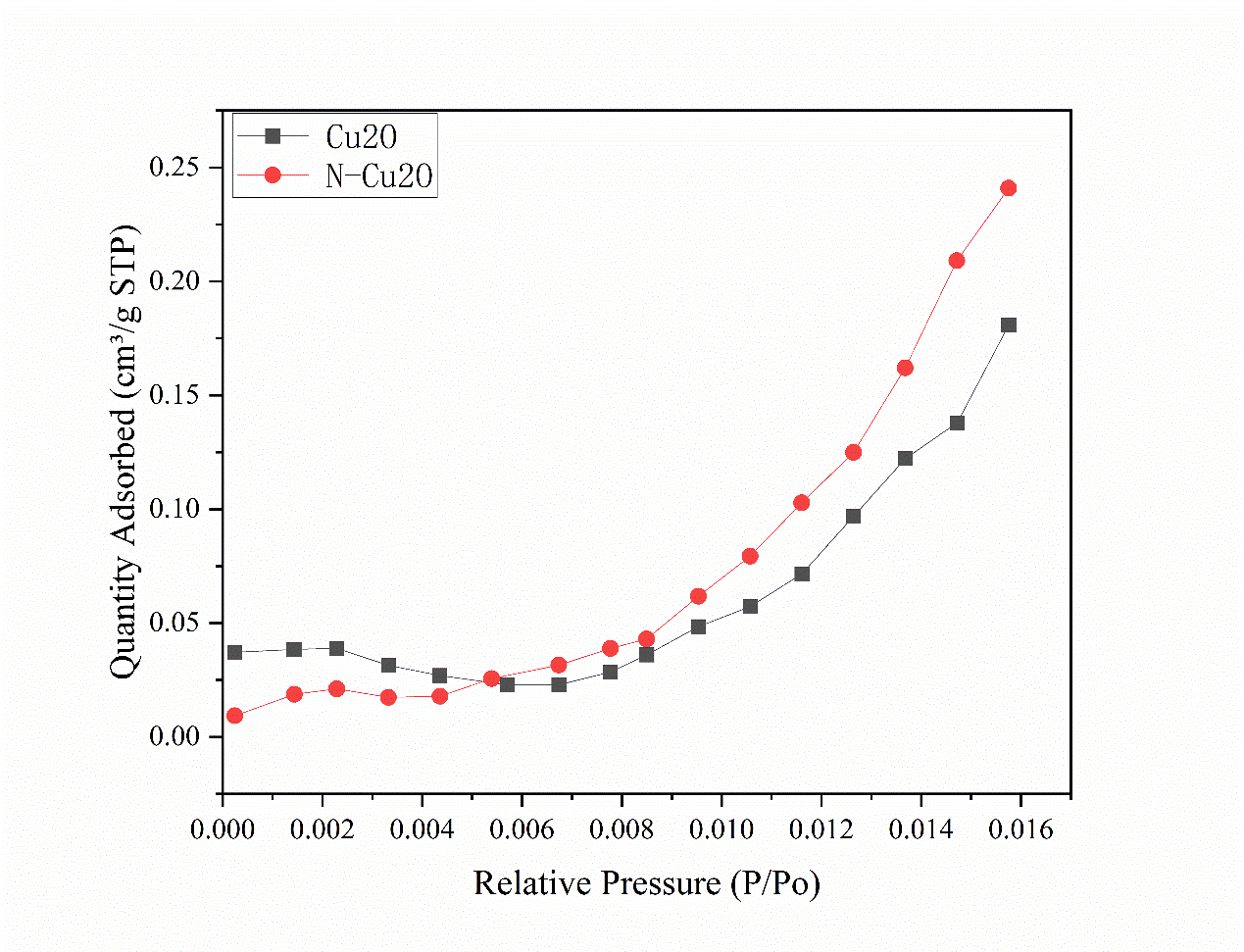}
\caption*{Figure S5. The BET adsorption isotherm curve of CO\textsubscript{2} for Cu\textsubscript{2}O and N-Cu\textsubscript{2}O}
\end{figure}

\begin{figure}[H]
    \centering
    \includegraphics[width=1\linewidth]{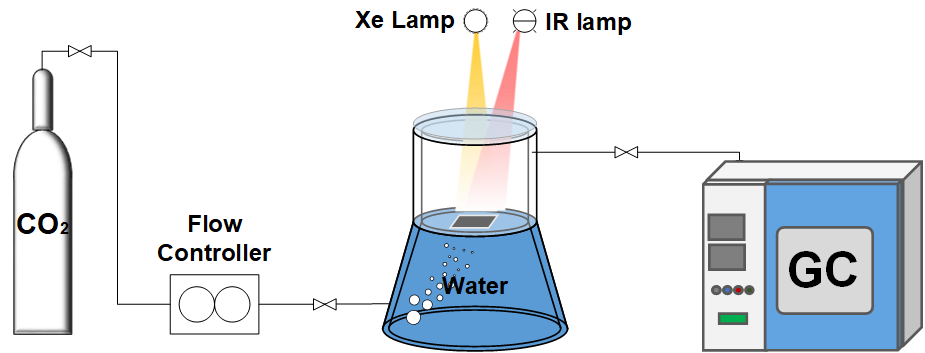}
Figure S6. Schematic illustration of the experimental setup for photocatalytic CO\textsubscript{2} reduction 
\end{figure}

\begin{figure}[H]
    \centering
    \includegraphics[width=1\linewidth]{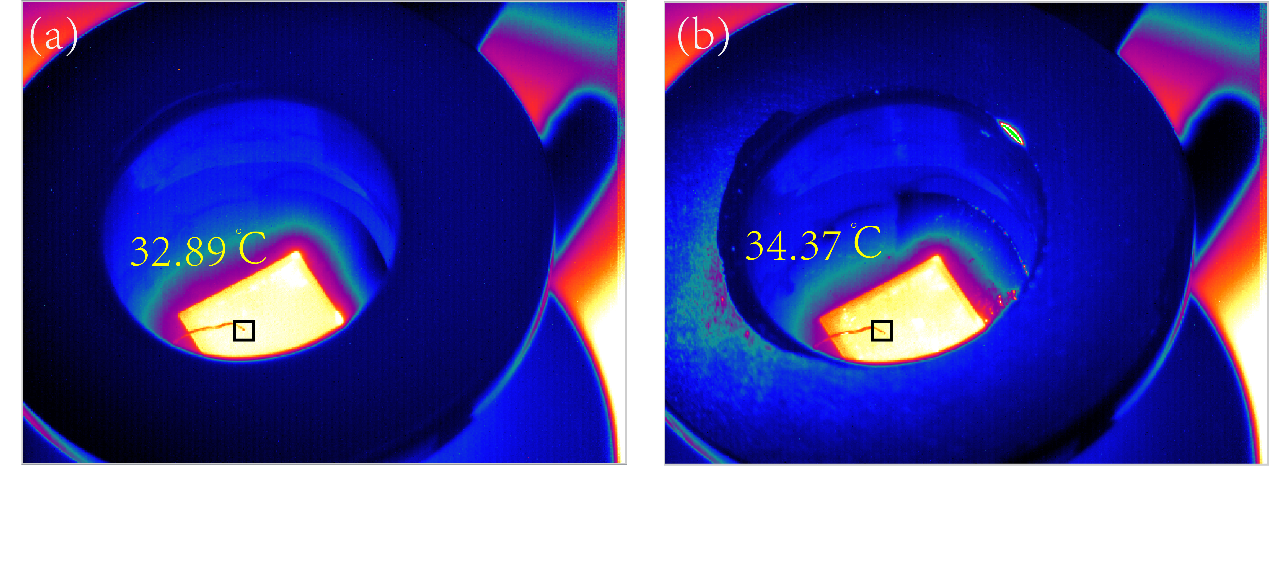}
Figure S7. Temperature of the catalyst surface under simultaneous Xe and IR lamp irradiation, measured by infrared thermography and a thermocouple
\end{figure}

\begin{figure}[H]
    \centering
    \includegraphics[width=1\linewidth]{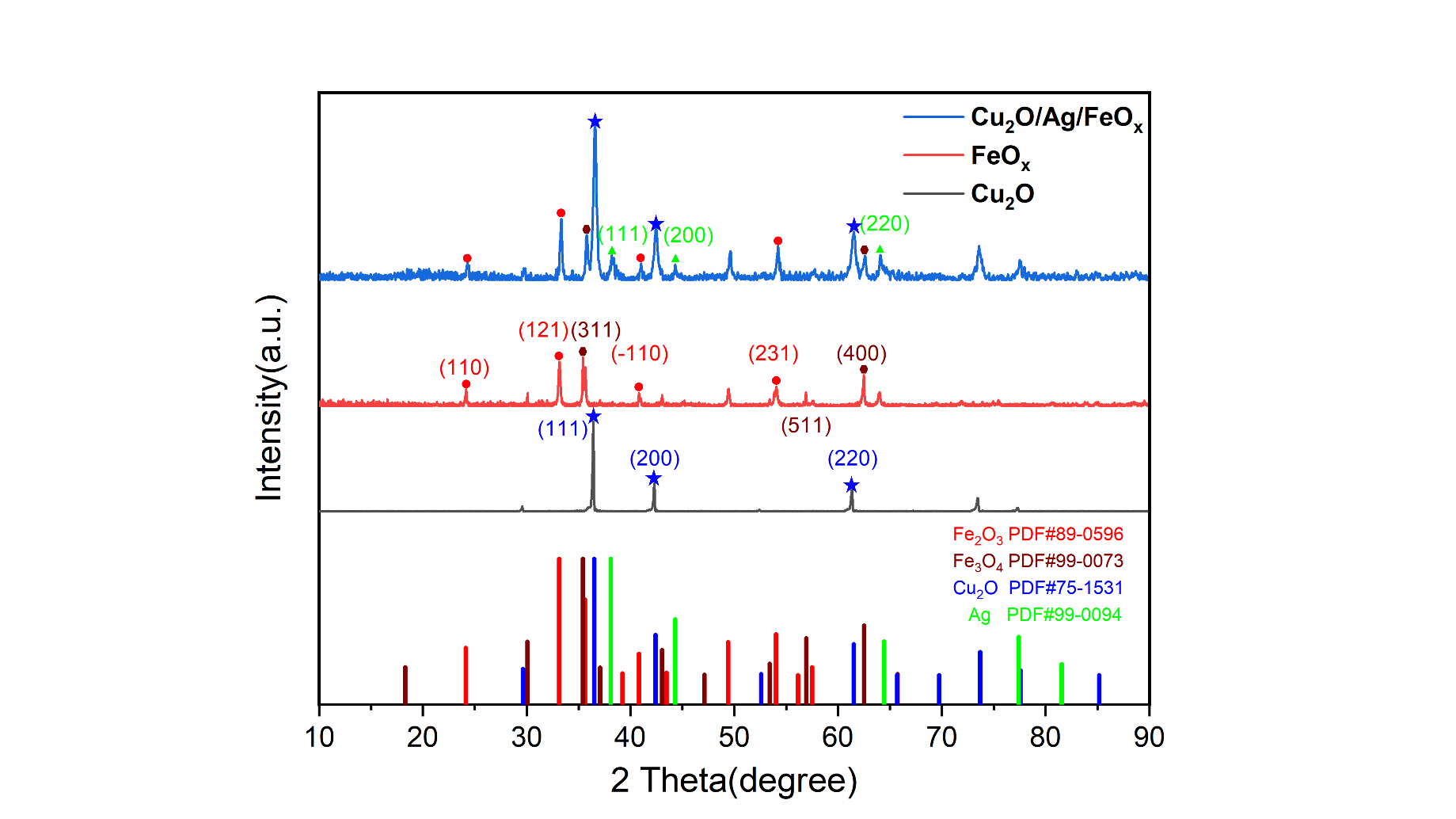}
Figure S8. XRD patterns of Cu\textsubscript{2}O/Ag/FeO\textsubscript{x},FeO\textsubscript{x} and Cu\textsubscript{2}O 
\end{figure}

\begin{figure}[H]
    \centering
    \includegraphics[width=1\linewidth]{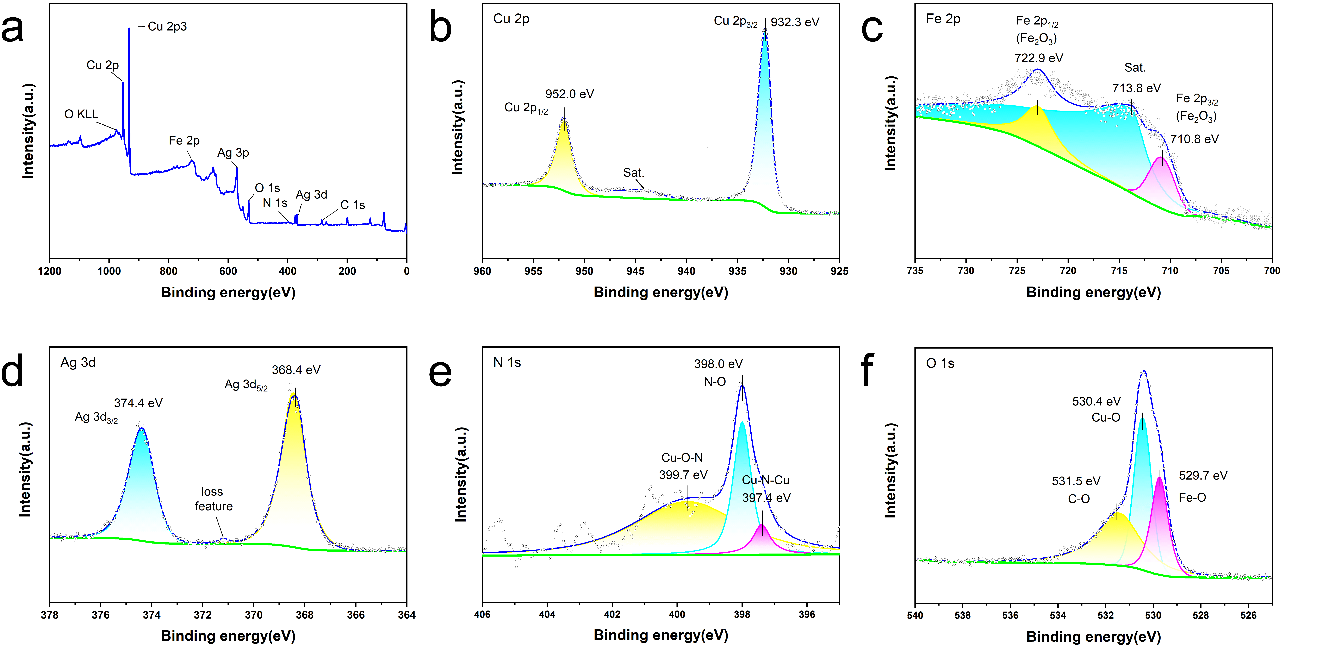}
Figure S9. XPS spectra of N-CAF : a) survey, high-resolution b) Cu 2p, c) Fe 2p, d) Ag 3d, e) Fe 2p and f) O 1s.
\end{figure}

\begin{figure}[H]
    \centering
    \includegraphics[width=1\linewidth]{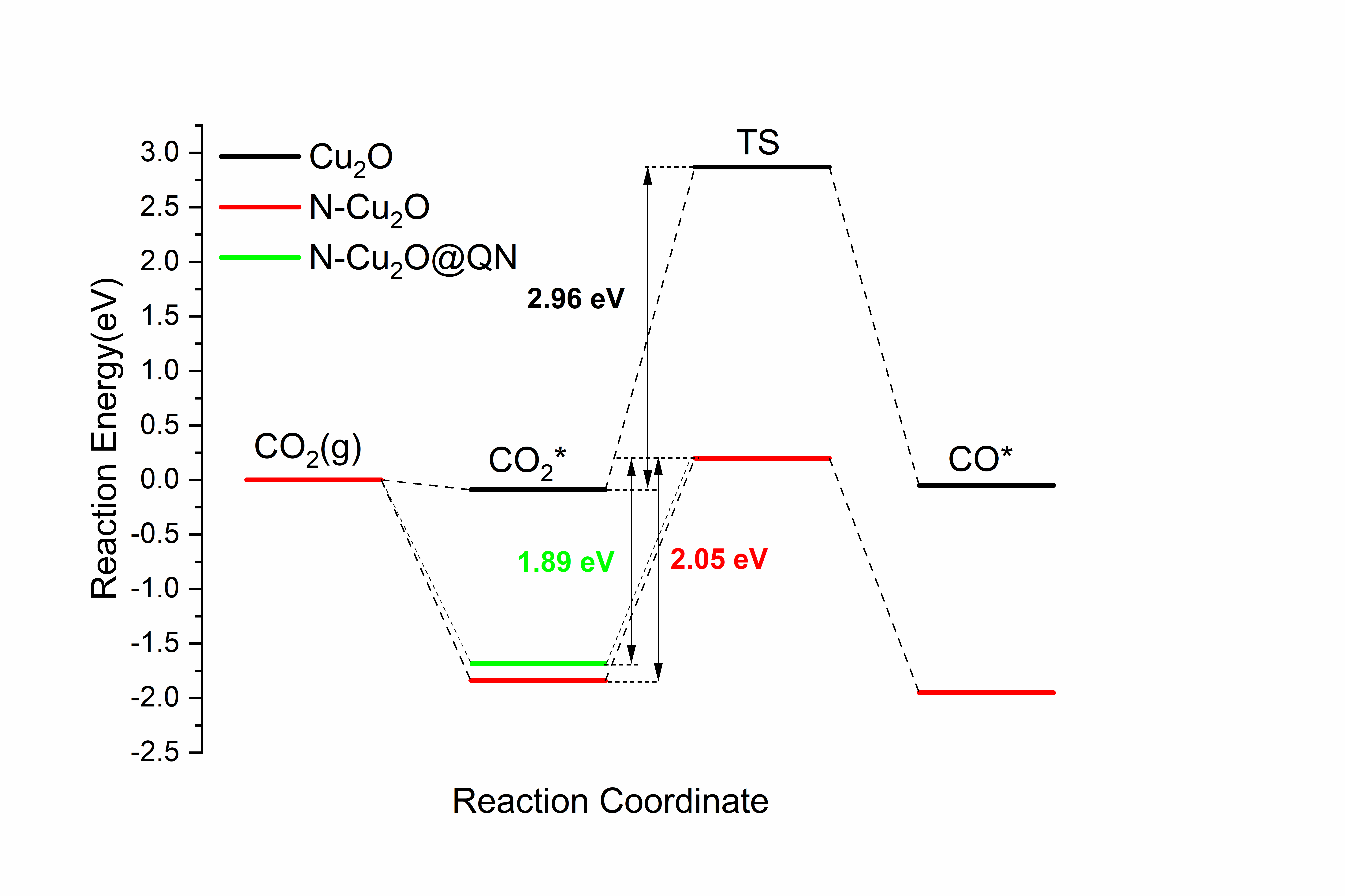}
Figure S10. The dissociation reaction pathways of CO\textsubscript{2} molecules on different catalyst surfaces and under resonant coupling conditions
\end{figure}
\textbf{
\subsection{S8 Table}
}

Table S1. Phonon energy of different vibration modes for CO\textsubscript{2} of gas state.
\begin{table}[H]
    \centering
    \begin{tabular}{|c|c|c|}
        \hline
        & Frequency cm$^{-1}$& Energy (meV)\\
        \hline
        1 & 2282.76 & 293.29 \\
        \hline
        2 & 1272.83 & 163.53 \\
        \hline
        3 & 611.39 & 78.55 \\
        \hline
        4 & 611.39 & 78.55 \\
        \hline
    \end{tabular}
\end{table}

Table S2. Phonon energy of different vibration modes for CO\textsubscript{2} adsorbed on N-Cu\textsubscript{2}O surface,
\begin{table}[H]
    \centering
    \begin{tabular}{|c|c|c|}
    \hline
    \textbf{} & \textbf{Frequency (cm$^{-1}$)} & \textbf{Energy (meV)} \\ \hline
    1 & 1599.31 & 205.48 \\ \hline
    2 & 1136.79 & 146.05 \\ \hline
    3 & 708.66  & 91.05  \\ \hline
    4 & 704.46  & 90.50  \\ \hline
    5 & 468.82  & 60.23  \\ \hline
    \end{tabular}
\end{table}

Table S3. Parameters of the coupled oscillators model for Figure 4b
\begin{table}[H]
    \centering
    \begin{tabular}{|c|c|c|c|c|c|c|}
        \hline
        $r_q$& $\omega_p$& $\gamma_p$& $\omega_m$& $\gamma_m$& g& C\\
        $(\mu \text{m})$& $(\text{cm}^{-1})$& $(\text{cm}^{-1})$& $(\text{cm}^{-1})$& $(\text{cm}^{-1})$& & \\\hline
        0.82 & 1160 & 23 & 1159 & 15 & 6.5 & 0.171 \\
        0.90 & 1154 & 24 & 1159.5 & 15 & 5.4 & 0.138 \\
        1.00 & 1149 & 24 & 1159.5 & 15 & 6.0 & 0.154 \\
        1.10 & 1143 & 24 & 1158 & 15 & 7 & 0.179 \\
        1.20 & 1137 & 24 & 1158 & 15 & 11 & 0.282 \\
        1.30 & 1130 & 24 & 1159.5 & 15 & 6.6 & 0.179 \\
        1.34 & 1128 & 23 & 1159.5 & 15 & 7.2 & 0.184 \\
        1.44 & 1120 & 24 & 1159 & 15 & 7.6 & 0.195 \\\hline
    \end{tabular}
\end{table}
Table S4. Summary of CO\textsubscript{2} photo-reduction to CO with water oxidation.
\begin{figure}[H]
    \centering
    \includegraphics[width=1\linewidth]{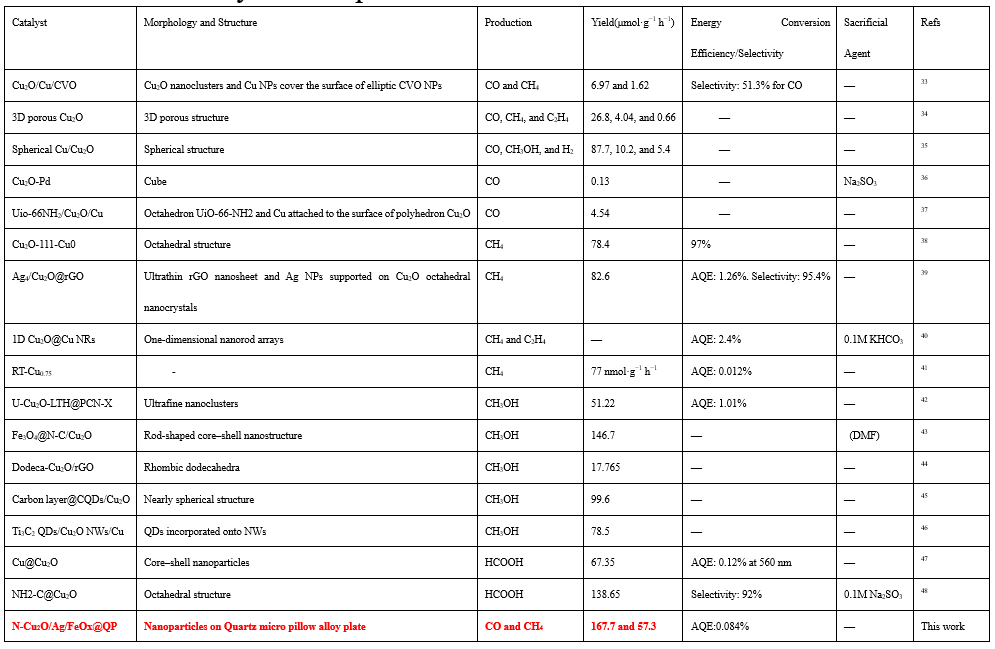}
\end{figure}
\textbf{
\subsection{References}
}

1 Solymosi, F. The bonding, structure and reactions of CO2 adsorbed on clean and promoted metal surfaces. \emph{Journal of Molecular Catalysis} \textbf{65}, 337-358, doi:10.1016/0304-5102(91)85070-i (1991).

2 Li, P. \emph{et al.} Hexahedron Prism-Anchored Octahedronal CeO2: Crystal Facet-Based Homojunction Promoting Efficient Solar Fuel Synthesis. \emph{J Am Chem Soc} \textbf{137}, 9547-9550, doi:10.1021/jacs.5b05926 (2015).

3 Neatu, S., Macia-Agullo, J. A. \& Garcia, H. Solar light photocatalytic CO2 reduction: general considerations and selected bench-mark photocatalysts. \emph{Int J Mol Sci} \textbf{15}, 5246-5262, doi:10.3390/ijms15045246 (2014).

4 Fujishima, A. \& Honda, K. Electrochemical photolysis of water at a semiconductor electrode. \emph{Nature} \textbf{238}, 37-38, doi:10.1038/238037a0 (1972).

5 Wu, H.-K., Li, Y.-H., Qi, M.-Y., Lin, Q. \& Xu, Y.-J. Enhanced photocatalytic CO2 reduction with suppressing H2 evolution via Pt cocatalyst and surface SiO2 coating. \emph{Applied Catalysis B: Environmental} \textbf{278}, doi:10.1016/j.apcatb.2020.119267 (2020).

6 Chang, X. \emph{et al.} Tuning Cu/Cu(2) O Interfaces for the Reduction of Carbon Dioxide to Methanol in Aqueous Solutions. \emph{Angew Chem Int Ed Engl} \textbf{57}, 15415-15419, doi:10.1002/anie.201805256 (2018).

7 Shi, X. \emph{et al.} Photoswitchable Chlorine Vacancies in Ultrathin Bi4O5Cl2 for Selective CO2 Photoreduction. \emph{ACS Catalysis} \textbf{12}, 3965-3973, doi:10.1021/acscatal.2c00157 (2022).

8 Wang, Y. G., Wiberg, K. B. \& Werstiuk, N. H. Correlation effects in EOM-CCSD for the excited states: evaluated by AIM localization index (LI) and delocalization index (DI). \emph{J Phys Chem A} \textbf{111}, 3592-3601, doi:10.1021/jp067579t (2007).

9 Zou, Y. \emph{et al.} Surface frustrated Lewis pairs in titanium nitride enable gas phase heterogeneous CO(2) photocatalysis. \emph{Nat Commun} \textbf{15}, 10604, doi:10.1038/s41467-024-54951-2 (2024).

10 Meryem, S. S., Nasreen, S., Siddique, M. \& Khan, R. An overview of the reaction conditions for an efficient photoconversion of CO2. \emph{Reviews in Chemical Engineering} \textbf{34}, 409-425, doi:10.1515/revce-2016-0016 (2018).

11 Lu, K. Q. \emph{et al.} Rationally designed transition metal hydroxide nanosheet arrays on graphene for artificial CO(2) reduction. \emph{Nat Commun} \textbf{11}, 5181, doi:10.1038/s41467-020-18944-1 (2020).

12 Li, H., Zhao, J., Luo, L., Du, J. \& Zeng, J. Symmetry-Breaking Sites for Activating Linear Carbon Dioxide Molecules. \emph{Acc Chem Res} \textbf{54}, 1454-1464, doi:10.1021/acs.accounts.0c00715 (2021).

13 Bols, M. L. \emph{et al.} Coordination and activation of nitrous oxide by iron zeolites. \emph{Nat Catal} \textbf{4}, 332-340,
doi:10.1038/s41929-021-00602-4 (2021).

14 Yu, Q. \& Bowman, J. M. Manipulating hydrogen bond dissociation rates and mechanisms in water dimer through vibrational strong coupling. \emph{Nat Commun} \textbf{14}, 3527, doi:10.1038/s41467-023-39212-y (2023).

15 Campos-Gonzalez-Angulo, J. A., Ribeiro, R. F. \& Yuen-Zhou, J. Resonant catalysis of thermally activated chemical reactions with vibrational polaritons. \emph{Nat Commun} \textbf{10}, 4685, doi:10.1038/s41467-019-12636-1 (2019).

16 Dolado, I. \emph{et al.} Remote near-field spectroscopy of vibrational strong coupling between organic molecules and phononic nanoresonators. \emph{Nat Commun} \textbf{13}, 6850, doi:10.1038/s41467-022-34393-4 (2022).

17 Tran, P. D., Wong, L. H., Barber, J. \& Loo, J. S. C. Recent advances in hybrid photocatalysts for solar fuel production. \emph{Energy \& Environmental Science} \textbf{5}, doi:10.1039/c2ee02849b (2012).

18 Christoforidis, K. C. \& Fornasiero, P. Photocatalysis for Hydrogen Production and CO2 Reduction: The Case of Copper‐Catalysts. \emph{ChemCatChem} \textbf{11}, 368-382, doi:10.1002/cctc.201801198 (2018).

19 Wu, Y. A. \emph{et al.} Facet-dependent active sites of a single Cu2O particle photocatalyst for CO2 reduction to methanol. \emph{Nature Energy} \textbf{4}, 957-968, doi:10.1038/s41560-019-0490-3 (2019).

20 Li, P. \emph{et al.} Reversible optical switching of highly confined phonon-polaritons with an ultrathin phase-change material. \emph{Nat Mater} \textbf{15}, 870-875, doi:10.1038/nmat4649 (2016).

21 Amarie, S. \& Keilmann, F. Erratum: Broadband-infrared assessment of phonon resonance in scattering-type near-field microscopy {[}Phys. Rev. B83, 045404 (2011){]}. \emph{Physical Review B} \textbf{84}, doi:10.1103/PhysRevB.84.199904 (2011).

22 Memmi, H., Benson, O., Sadofev, S. \& Kalusniak, S. Strong Coupling between Surface Plasmon Polaritons and Molecular Vibrations. \emph{Phys Rev Lett} \textbf{118}, 126802, doi:10.1103/PhysRevLett.118.126802 (2017).

23 Wu, R., Zhang, J., Shi, Y., Liu, D. \& Zhang, B. Metallic WO2-Carbon Mesoporous Nanowires as Highly Efficient Electrocatalysts for Hydrogen Evolution Reaction. \emph{J Am Chem Soc} \textbf{137}, 6983-6986, doi:10.1021/jacs.5b01330 (2015).

24 Malerba, C. \emph{et al.} Corrigendum to ``Absorption coefficient of bulk and thin film Cu2O'' {[}Sol. Energy Mater. Sol. Cells 95(10) (2011) 2848--2854{]}. \emph{Solar Energy Materials and Solar Cells} \textbf{98}, doi:10.1016/j.solmat.2011.10.018 (2012).

25 Zhang, W. \emph{et al.} Photocatalytic degradation mechanism of gaseous styrene over Au/TiO2@CNTs: Relevance of superficial state with deactivation mechanism. \emph{Applied Catalysis B: Environmental} \textbf{272}, doi:10.1016/j.apcatb.2020.118969 (2020).

26 Soltani, T. \emph{et al.} Effect of transition metal oxide cocatalyst on the photocatalytic activity of Ag loaded CaTiO3 for CO2 reduction with water and water splitting. \emph{Applied Catalysis B: Environmental} \textbf{286}, doi:10.1016/j.apcatb.2021.119899 (2021).

27 Perdew, J. P., Burke, K. \& Ernzerhof, M. Generalized gradient approximation made simple. \emph{Physical Review Letters} \textbf{77}, 3865-3868, doi:DOI 10.1103/PhysRevLett.77.3865 (1996).

28 Kresse, G. J., D. From Ultrasoft Pseudopotentials to the Projector Augmented-Wave Method. \emph{Phys. Rev. B: Condens. Matter Mater. Phys} (1994).

29 Blochl, P. E. Projector augmented-wave method. \emph{Phys Rev B Condens Matter} \textbf{50}, 17953-17979, doi:10.1103/physrevb.50.17953 (1994).

30 Sheppard, D., Xiao, P., Chemelewski, W., Johnson, D. D. \& Henkelman, G. A generalized solid-state nudged elastic band method. \emph{J Chem Phys} \textbf{136}, 074103, doi:10.1063/1.3684549 (2012).

31 Henkelman, R. M., Stanisz, G. J. \& Graham, S. J. A multicenter measurement of magnetization transfer ratio in normal white matter. \emph{J Magn Reson Imaging} \textbf{11}, 568, doi:10.1002/(sici)1522-2586(200005)11:5\textless568::aid-jmri14\textgreater3.0.co;2-8 (2000).

32 Amarie, S. \& Keilmann, F. Broadband-infrared assessment of phonon resonance in scattering-type near-field microscopy. \emph{Physical Review B} \textbf{83}, doi:10.1103/PhysRevB.83.045404 (2011).

33 Song, Y. \emph{et al.} Z-Scheme Cu2O/Cu/Cu3V2O7(OH)2·2H2O Heterostructures for Efficient Visible-Light Photocatalytic CO2 Reduction. \emph{ACS Applied Energy Materials} \textbf{5}, 10542-10552, doi:10.1021/acsaem.2c01252 (2022).

34 Cui, L. \emph{et al.} Three-dimensional porous Cu2O with dendrite for efficient photocatalytic reduction of CO2 under visible light. \emph{Applied Surface Science} \textbf{581}, doi:10.1016/j.apsusc.2021.152343 (2022).

35 Zheng, Y. \emph{et al.} Shape-Dependent Performance of Cu/Cu(2) O for Photocatalytic Reduction of CO(2). \emph{ChemSusChem} \textbf{15}, e202200216, doi:10.1002/cssc.202200216 (2022).

36 Zhang, X. \emph{et al.} Palladium-modified cuprous(i) oxide with 100 facets for photocatalytic CO(2) reduction. \emph{Nanoscale} \textbf{13}, 2883-2890, doi:10.1039/d0nr07703h (2021).

37 Zhao, X. \emph{et al.} Cu media constructed Z-scheme heterojunction of UiO-66-NH2/Cu2O/Cu for enhanced photocatalytic induction of CO2. \emph{Applied Surface Science} \textbf{545}, doi:10.1016/j.apsusc.2021.148967 (2021).

38 Deng, Y. \emph{et al.} Synergy Effect between Facet and Zero-Valent Copper for Selectivity Photocatalytic Methane Formation from CO2. \emph{ACS Catalysis} \textbf{12}, 4526-4533, doi:10.1021/acscatal.2c00167 (2022).

39 Tang, Z. \emph{et al.} Ternary heterojunction in rGO-coated Ag/Cu2O catalysts for boosting selective photocatalytic CO2 reduction into CH4. \emph{Applied Catalysis B: Environmental} \textbf{311}, doi:10.1016/j.apcatb.2022.121371 (2022).

40 Zhou, J. \emph{et al.} Facile in situ fabrication of Cu2O@Cu metal-semiconductor heterostructured nanorods for efficient visible-light driven CO2 reduction. \emph{Chemical Engineering Journal} \textbf{385}, doi:10.1016/j.cej.2019.123940 (2020).

41 Ali, S. \emph{et al.} Sustained, photocatalytic CO2 reduction to CH4 in a continuous flow reactor by earth-abundant materials: Reduced titania-Cu2O Z-scheme heterostructures. \emph{Applied Catalysis B: Environmental} \textbf{279}, doi:10.1016/j.apcatb.2020.119344 (2020).

42 Yao, S. \emph{et al.} Anchoring ultrafine Cu2O nanocluster on PCN for CO2 photoreduction in water vapor with much improved stability. \emph{Applied Catalysis B: Environmental} \textbf{317}, doi:10.1016/j.apcatb.2022.121702 (2022).

43 Kazemi Movahed, S., Najinasab, A., Nikbakht, R. \& Dabiri, M. Visible light assisted photocatalytic reduction of CO2 to methanol using Fe3O4@N-C/Cu2O nanostructure photocatalyst. \emph{Journal of Photochemistry and Photobiology A: Chemistry} \textbf{401}, doi:10.1016/j.jphotochem.2020.112763 (2020).

44 Liu, S.-H., Lu, J.-S., Pu, Y.-C. \& Fan, H.-C. Enhanced photoreduction of CO2 into methanol by facet-dependent Cu2O/reduce graphene oxide. \emph{Journal of CO2 Utilization} \textbf{33}, 171-178, doi:10.1016/j.jcou.2019.05.020 (2019).

45 Li, H. \emph{et al.} Carbon quantum dots and carbon layer double protected cuprous oxide for efficient visible light CO(2) reduction. \emph{Chem Commun (Camb)} \textbf{55}, 4419-4422, doi:10.1039/c9cc00830f (2019).

46 Zeng, Z. \emph{et al.} Boosting the Photocatalytic Ability of Cu2O Nanowires for CO2 Conversion by MXene Quantum Dots. \emph{Advanced Functional Materials} \textbf{29}, doi:10.1002/adfm.201806500 (2018).

47 Wang, H. \emph{et al.} Photocatalytic CO2 reduction to HCOOH over core-shell Cu@Cu2O catalysts. \emph{Catalysis Communications} \textbf{162}, doi:10.1016/j.catcom.2021.106372 (2022).

48 Zhu, Q. \emph{et al.} CO2 reduction to formic acid via NH2-C@Cu2O photocatalyst in situ derived from amino modified Cu-MOF. \emph{Journal of CO2 Utilization} \textbf{54}, doi:10.1016/j.jcou.2021.101781 (2021).

\end{document}